\documentclass{aa}  

\usepackage{graphicx}
\usepackage{multirow}
\usepackage{arydshln}
\usepackage{adjustbox}

\usepackage{txfonts}

\usepackage{hyperref}
\hypersetup{
    linkcolor=blue,
    citecolor=blue,
    colorlinks=true,
    filecolor=blue,      
    urlcolor=blue}

\begin{document} 

   \title{Gas motion in the intracluster
medium of the Virgo cluster replica}

   \titlerunning{Gas motion in Virgo's ICM}

    \author{Th\'eo Lebeau\inst{1} \thanks{Corresponding author: \href{mailto:theo.lebeau@universite-paris-saclay.fr}{theo.lebeau@universite-paris-saclay.fr}}, Stefano Ettori\inst{2,3,1}, Jenny G. Sorce\inst{4,1}, Nabila Aghanim\inst{1}, Jade Pasté\inst{1}}

   \authorrunning{Lebeau et al.}

   \institute{Université Paris-Saclay, CNRS, Institut d’Astrophysique Spatiale, 91405 Orsay, France
        \and
            INAF, Osservatorio di Astrofisica e Scienza dello Spazio, via Piero Gobetti 93/3, 40129 Bologna, Italy
        \and
          INFN, Sezione di Bologna, viale Berti Pichat 6/2, 40127 Bologna, Italy
        \and
             Université de Lille, CNRS, Centrale Lille, UMR 9189 CRIStAL, F-59000 Lille, France}

   \date{Received 17 June 2025 / Accepted 19 January 2026 }
 
  \abstract 
  {Within the deep gravitational potential of galaxy clusters lies the intracluster medium (ICM). At the first order, it is considered to be at hydrostatic equilibrium within the potential well. However, evidence is growing on the observational side and in numerical simulations that the ICM dynamics is non-negligible from an energetic point of view, is mostly turbulent in origin, and provides non-thermal pressure support to the equilibrium. In this work, we intend to characterise the properties of the velocity field in the ICM of a simulated replica of the Virgo cluster. We first studied the 3D and projected properties of the ICM velocity field by computing its probability density functions (PDFs) and its statistical moments. We then estimated the non-thermal pressure fraction from an effective turbulent Mach number, including the velocity dispersion. We finally computed the velocity structure function (VSF) from projected maps of the sightline velocity. In this paper, we first show that the components of the 3D velocity field and the projected quantities along equivalent sightlines are anisotropic and affected by the accretion of gas from filaments. Then, we compare the mean statistical moments of the 3D velocity field to the mean properties of 100 random projections. We show, in particular, an almost linear relation between the standard deviation estimated from direct simulation outputs and sightline velocity dispersion projections, which are comparable to the line broadening of X-ray atomic lines. However, this linear relation does not hold between the direct simulation outputs and the standard deviation of the sightline velocity projections, which are comparable to the line shift of X-ray atomic lines. We find a non-thermal pressure fraction of around $6\%$ within $R_{500}$ and $9\%$ within $R_{vir}$ from sightline velocity dispersion, which is in good agreement with direct simulation outputs. Finally, we show that the VSF might probe the turbulent injection scale of active galactic nucleus (AGN) feedback.}


   \maketitle

\section{Introduction}

Initial over-densities accreted matter through cosmic history and formed the current clusters of galaxies, which nowadays constitute the nodes of the cosmic web. Simultaneously, cosmic gas fell into these deep gravitational potential wells and was thus gravitationally heated up to $10^8$~K, forming the intracluster medium (ICM; see \citeauthor{kravtsov2012formation} \citeyear{kravtsov2012formation} for a review). In this context, gravity is considered to be by far the most dominant process in the dynamics of the ICM; we thus usually consider that pressure only comes from gravitational heating, inducing purely thermal pressure. Consequently, the ICM gas is assumed to be at hydrostatic equilibrium, allowing us to derive the mass of clusters from the measure of the ICM pressure via the thermal Sunyaev-Zel'dovich effect (tSZ, \citeauthor{sunyaev1972observations} \citeyear{sunyaev1972observations}) in the sub-millimetre wavelengths and the measure of the ICM density and temperature via bremsstrahlung \citep[e.g.][]{sarazin1986x,ettori2013mass} in the X-rays \citep[see e.g.][for mass estimation of clusters and their use as a cosmological probes]{salvati2018constraints,pratt2019galaxy,2024A&A...690A.238Aymerich}. \\

The main source of deviation from hydrostatic equilibrium could be gas dynamics in the ICM \citep[e.g.][]{rasia2004dynamical}, which could contribute significantly, in addition to thermal pressure, to the total pressure support \citep[][]{nelson2014hydrodynamic}. In observations, properties of the velocity field in the ICM can be probed with X-ray spectroscopy, provided there is a sufficient resolution to measure precise line shift and broadening to derive sightline velocity and velocity dispersion (see e.g. \citeauthor{2012MNRAS.422.2712Zhuravleva} \citeyear{2012MNRAS.422.2712Zhuravleva}, and \citeauthor{2019SSRv..215...24Simionescu} \citeyear{2019SSRv..215...24Simionescu} for a review). On the one hand, the XRISM telescope \citep{xrism2020science}, replacing the Hitomi telescope \citep[e.g.][]{hitomi2016quiescent,hitomi2018atmospheric}, resolves line shift and broadening, although with a limited spatial resolution with respect to XMM-Newton or Chandra. For instance, low, non-thermal pressure has been measured in the core of the A2029 cluster \citep[][]{2025ApJ...982L...5XRISM} and in its outskirts \citep[][]{2025PASJ...77S.242XRISM}, but also in Coma \citep{2025ApJ...985L..20XRISM-Coma} and other clusters \citep[all compared in][]{2025ApJ...993L..11XRISM_comp}. On the other hand, a refined analysis of XMM-Newton data permitted the mapping of the velocity field in nearby clusters \citep[e.g.][]{2022MNRAS.513.1932Gatuzz_Centaurus,gatuzz2023measuring,2023MNRAS.522.2325Gatuzz_Ophiuchus,2024A&A...692A.108Gatuzz_A3266}, including the Virgo cluster \citep[][]{2022MNRAS.511.4511Gatuzz_Virgo}. It allowed the study of statistical properties of the velocity field through the velocity structure function (VSF), although with large uncertainties in the measurement of the velocity due to its lower spectral resolution than XRISM. In the future, the NewAthena telescope \citep[][]{2025NatAs...9...36Cruise_NewAthena} will allow both high spatial and spectral resolution.\\

The quantification of the velocity field properties from observations still relies on simplified assumptions: isotropy, homogeneity, and Gaussianity \citep[see e.g.][]{gatuzz2023measuring}. Moreover, turbulence in the core of clusters is not fully understood yet, even though it is a very active research field thanks to simulations \citep[e.g.][]{norman1999cluster,nagai2007effects,gaspari2013constraining,porter2015vorticity,vazza2017turbulence,valles2021troubled,2024A&A...690A..20Ayromlou,2025A&A...693A.263Groth,2026A&A...705A.129Vazza}. In particular, the role of accretion and merging shocks \citep[e.g.][]{zuhone2011parameter,iapichino2011turbulence,nagai2013predicting}, active galactic nucleus feedback (AGN) \citep[e.g.][]{vazza2013thermal,bourne2017agn,2025A&A...697A.138Sotira}, stratification \citep[e.g.][]{shi2018multiscale,mohapatra2020turbulence,simonte2022exploring,wang2023turbulent}, and multiple phases \citep[e.g.][]{wang2021non,mohapatra2022characterizing} have been studied. From an observational standpoint, indirectly quantifying turbulence from density fluctuations was proposed \citep[e.g.][]{gaspari2013constraining,2014ApJ...788L..13Zhuravleva}; this can be derived from X-ray surface brightness fluctuations \citep[e.g.][]{2012MNRAS.421.1123C_Churazov,dupourque2023investigating,dupourque2024chex,2024MNRAS.528.7274Heinrich}, pressure fluctuations \citep[e.g.][]{schuecker2004probing,khatri2016thermal,romero2023inf,romero2024surface,2025A&A...694A.182Adam}, or temperature fluctuations \citep[e.g.][]{2024A&A...682A..45Lovisari}, and thus completes spectroscopic studies. This work therefore aims to thoroughly study the properties of the velocity field in the ICM through simulations, using both 3D direct simulation outputs and projections of the sightline velocity and sightline velocity dispersion. \\

We took advantage of a state-of-the-art cosmological hydrodynamical simulated replica of the Virgo cluster in its local environment \citep{sorce2021hydrodynamical} to conduct this study, allowing direct comparison with observations. We first describe the simulation and present projections of the sightline velocity and sightline velocity dispersion in Sect. \ref{sec:2 methodo}. We study the statistical properties of the 3D velocity field components and projected sightline velocities, via the probability distribution function (PDF) and its statistical moments, in Sect. \ref{sec:3 stat}. We then derive the turbulent Mach number and the non-thermal pressure support from 3D and projected quantities in Sect. \ref{sec:4 Mach}. We finally compute the velocity structure function (VSF) in Sect. \ref{sec:VSF} before discussing the results in Sect. \ref{discussion} and concluding in Sect. \ref{conclusion}. 

\begin{figure*}
    \centering
    \includegraphics[trim= 0 0 50 0,clip, width=0.9\linewidth]{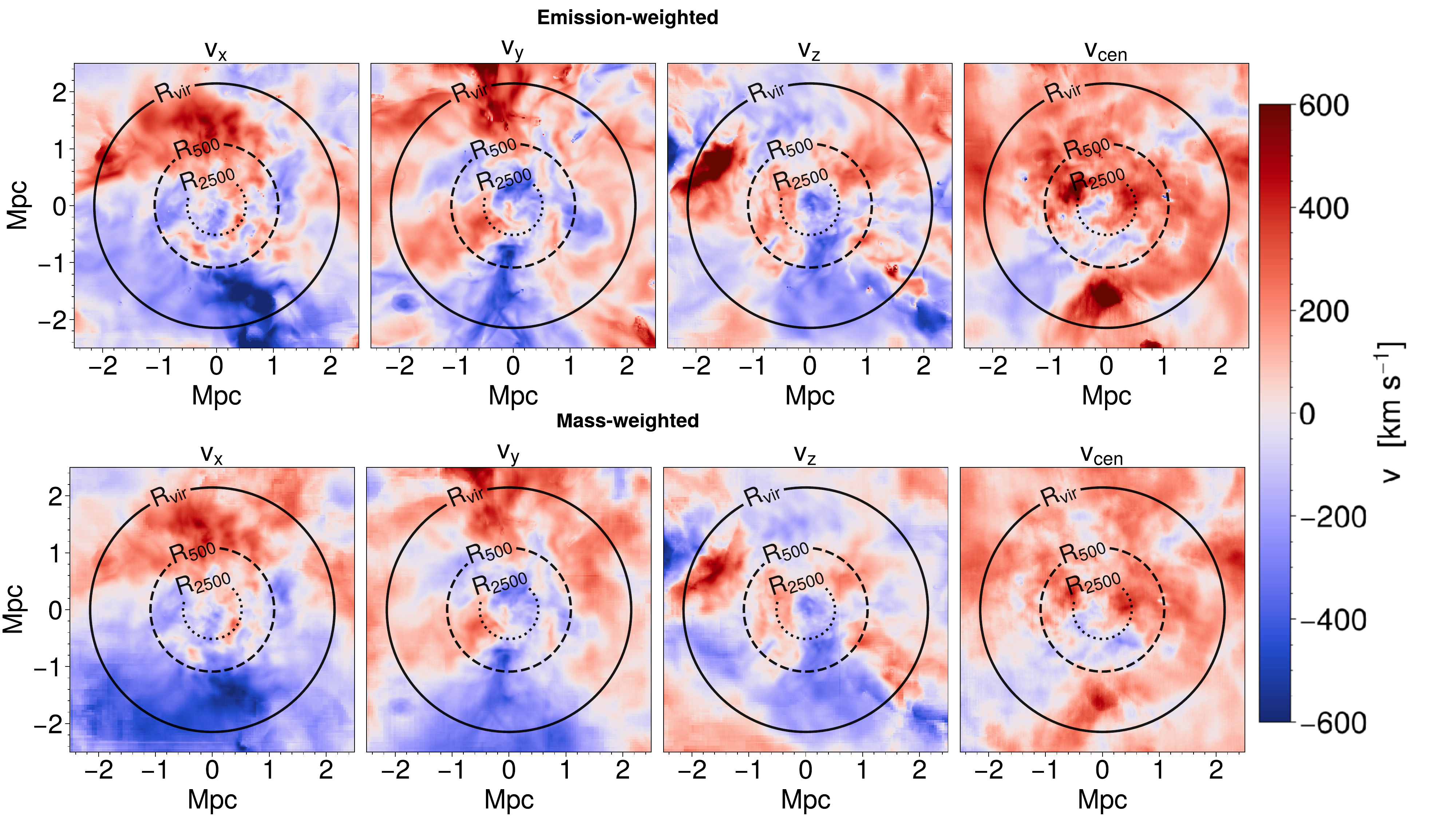}
    \caption{Projections of the velocity integrated along the $v_x$,$v_y$,$v_z$ and $v_{cen}$ sightlines from left to right. The top (bottom) row presents the EW (MW) maps. The solid, dashed and dotted black circles respectively represent $R_{vir}$, $R_{500}$ and $R_{2500}$. The colour bar goes from $-600~\mathrm{km\,s^{-1}}$ (blue), meaning an average gas flow coming from the foreground of Virgo and going away from the observer, to $600~\mathrm{km\,s^{-1}}$ (red), meaning, inversely, an average gas flow coming from the background of Virgo and coming toward the observer.}
    \label{los vel maps}
\end{figure*}

\section{Methodology}
\label{sec:2 methodo}

\subsection{The Virgo replica simulation}

We succinctly present the Virgo replica simulation already used in \citet{2024A&A...682A.157Lebeau,2024A&A...689A..19Lebeau,2025A&A...704A..14Lebeau}; for more details, see \cite{sorce2021hydrodynamical}. Contrary to previous simulations attempting to reproduce the Virgo cluster \citep[e.g.][]{hoffman1980dynamical,li2014modeling,moran2014globular, zhu2014next}, this replica accounts for its local environment. To create it, the initial conditions of this simulation were obtained by reconstructing the initial density field of the local Universe by applying a reverse Zel'dovich approximation \citep{2013MNRAS.430..888Doumler} to the position and peculiar velocities of galaxies in the Cosmicflows-2 \citep[][]{tully2013cosmicflows} dataset (see \citeauthor{sorce2016cosmicflows} \citeyear{sorce2016cosmicflows} for details). \cite{sorce2019virgo} then increased the resolution in the region around Virgo and simulated 200 dark matter (DM) replicas of this cluster within its local environment. Among the realisations, the most representative of the 200, in terms of merging history and average properties regarding the full sample, was reused to run a zoom-in, high-resolution hydrodynamical counterpart \citep[][]{sorce2021hydrodynamical}. Both the DM-only and the hydrodynamical run show a very good agreement with observations (see more details in \citeauthor{sorce2021hydrodynamical} \citeyear{sorce2021hydrodynamical} and \citeauthor{2024A&A...682A.157Lebeau} \citeyear{2024A&A...682A.157Lebeau}), they notably reproduce Virgo's local environment well, including the filaments along our observer's sightline and the group of galaxies falling onto Virgo. It is worth noting that the initial conditions used to produce this Virgo replica have been reused in the joint LOCALIZATION\footnote{\url{https://localization.ias.universite-paris-saclay.fr/}}/SLOW\footnote{\url{https://www.usm.lmu.de/~dolag/Simulations/\#SLOW}} project \citep[see e.g.][]{dolag2023simulating,2024A&A...687A.253HernandezSLOWII,2025arXiv250715858HernandezSLOWV}. In reproducing the initial conditions of the local Universe, this so-called backward approach differs from the Bayesian forward approach, used in other projects such as ELUCID \citep{2014ApJ...794...94WangELUCID} or SIBELIUS \citep{2022MNRAS.509.1432SawalaSIBELIUS}, where the initial density field is inferred from the observed density field through a Hamiltonian Markov chain method (see Section 3.2 of \citeauthor{lebeau2025impact} \citeyear{lebeau2025impact} for more details and references). \\
 
This simulation uses the \cite{planckcosmoparam2014} cosmological parameters; that is, a Hubble constant of $H_0=67.77~\mathrm{km~s^{-1}~Mpc^{-1}}$, a dark-energy density of $\Omega_{\Lambda}=0.693$, a total matter density of $\Omega_\mathrm{m}=0.307$, a baryonic density of $\Omega_\mathrm{b}=0.048$, an amplitude of the matter power spectrum at 8~$\mathrm{Mpc~h^{-1}}$, $\sigma_8 = 0.829$,  and a spectral index of $n_\mathrm{s}=0.961$. The adaptive mesh refinement (AMR) hydrodynamical code {\tt RAMSES} \citep{teyssier2002cosmological} was used to produce this simulation. Within the 500~$\mathrm{Mpc~h^{-1}}$ local Universe box, the zoom region is a 30~Mpc diameter sphere with an effective resolution of $8192^3$ DM particles with a mass of $m_{\mathrm{DM}}=3\times10^7~\mathrm{M_\odot}$. The finest cell size of the AMR grid is 0.35~kpc. The simulation contains additional sub-grid models for radiative gas cooling and heating, star formation, and kinetic feedback from a type II supernova (SN) and AGN, similar to the Horizon-AGN implementation of \cite{dubois2014dancing,dubois2016horizon}. Moreover, we improved the AGN feedback model by orienting the jet according to the black-hole spin (see \citeauthor{dubois2021introducing} \citeyear{dubois2021introducing} for details).\\

\subsection{Projected velocity maps}

The {\tt rdramses} code\footnote{\url{https://github.com/florentrenaud/rdramses}, first used in \cite{renaud2013sub}.} was used to extract the DM particles and gas cell properties from the simulation. The Virgo DM halo and its galaxies were identified by applying the \cite{tweed2009building} halo finder, respectively, to the DM particles and the star particles. In this work, we define the gas as the baryonic component in the simulation cells \citep[as in][]{2024A&A...689A..19Lebeau}, in which hydrodynamics equations in the Eulerian formalism are solved; it thus does not include the stars. It is important to note that, for this study, we extracted gas cells up to a resolution of 22.5~kpc at best, meaning that we did not go deeper than this resolution level in the cell tree of the simulation. This choice is motivated by the fact that we will hardly be able to reach a spatial resolution below 20~kpc for most of the clusters observed in the X-rays or sub-millimetre wavelengths, and that below this scale complex microphysics processes become important. In addition, we found almost identical velocity structure functions with twice as many resolved maps (see right panel of Fig. \ref{app: val range fig + vsf comp 222 444}), showing that the chosen resolution is sufficient for the studied gas dynamics. In addition, we restricted our study to ICM cells by applying a temperature cut at $T=10^7$~K (see phase diagram of cosmic gas in e.g. \citeauthor{2019MNRAS.486.3766Martizzi} \citeyear{2019MNRAS.486.3766Martizzi},\citeauthor{gouin2022gas} \citeyear{gouin2022gas}). We produced projections of the velocity fields following \citet{2024A&A...689A..19Lebeau,2024A&A...682A.157Lebeau,2025A&A...704A..14Lebeau}. To mimic what could be derived from X-ray spectra atomic line shifts, we computed the mean velocity of ICM simulation cells along a given sightline in the Virgo rest frame. In other words, the velocity of the DM halo, as found by the halo finder, is subtracted from each cell. This procedure is similar to choosing a reference wavelength, typically that of the brightest cluster galaxy (BCG), to estimate the Doppler shift. The sightline velocity is thus defined as 

\begin{equation}
    v_\mathrm{q} = \frac{\sum_\mathrm{i} q_\mathrm{i}v_\mathrm{i}}{\sum_\mathrm{i}q_\mathrm{i}}
,\end{equation}

\noindent with $q_\mathrm{i}$ being the weight. We computed both mass-weighted (MW) projections, i.e. $q_\mathrm{i}=m_\mathrm{i}$, the cell mass, and emission-weighted (EW) projections, i.e. $q_\mathrm{i}=\rho_\mathrm{i}^2$, which is the squared cell density, as in \cite{roncarelli2018athena}. We also mimicked the X-ray spectra's atomic line broadening by integrating the velocity dispersion along the sightline defined as

\begin{equation}
    \sigma_\mathrm{q}^2 = \frac{\sum_\mathrm{i} q_\mathrm{i}v_\mathrm{i}^2}{\sum_\mathrm{i}q_\mathrm{i}}-v_\mathrm{q}^2
    \label{equ:sigma}
,\end{equation}

\noindent with $q_\mathrm{i}$ being the same weightings as for the integrated sightline velocity. \\

The sightline velocity and velocity dispersion are presented in Fig. \ref{los vel maps} and Fig. \ref{los vel disp maps} of Appendix \ref{app: vel disp los projs}. From left to right, we show the projections along the $x$, $y,$ and $z$ simulation box axes. We also show the projection along the $cen$ sightline \citep[as in][]{2024A&A...689A..19Lebeau,2024A&A...682A.157Lebeau}, that is, between Virgo and the centre of the local Universe simulation box in which the Virgo zoom-in region is simulated. It is equivalent to our own observer's sightline. For both figures, the top row presents the EW projections, and the bottom row presents the MW projections. The projections are $5$~Mpc wide and contain $222^2$ pixels of $22.5$~kpc in width. \\

The sightline velocity projections (see Fig. \ref{los vel maps}) were built so that the sightline points towards the observer; thus, the redder (the bluer) the pixels, the more positive (negative) the integrated velocity is, meaning that, on average, the gas falls onto Virgo from the background (foreground) in this given sightline. The colour convention follows that of other works on projected sightline velocity of the ICM gas \citep[e.g.][]{gatuzz2023measuring}, but it is reversed compared to that usually used in studies of the sightline velocities of galaxies associating red (blue) with redshift (blueshift). We observe that, whatever the weighting method, the $v_\mathrm{x}$ projection is bimodal with a positive (negative) velocity in the upper (lower) part of the map; this is because Virgo is connected to two filaments feeding gas to Virgo and located in the $\{x<0,y>0\}$ ($\{x>0,y<0\}$) area of the simulation box (see Fig. 1 of \citeauthor{2025A&A...704A..14Lebeau} \citeyear{2025A&A...704A..14Lebeau}). The $v_y$ and $v_\mathrm{z}$ are perpendicular to the filaments, so their projections present a spatially scattered distribution of positive and negative velocities. The $v_\mathrm{cen}$ projection is dominated by positive velocities given that it is almost aligned with filaments and that the matter flow in the background filament dominates that of the foreground one (see also Fig. 1 of \citeauthor{2025A&A...704A..14Lebeau} \citeyear{2025A&A...704A..14Lebeau}). We also note that the EW maps are more contrasted than their MW counterparts; the former enhances the denser regions when the latter shows a smoother distribution, which induces differences in the statistical-moment values of their respective velocity fields, as we show later. \\

\begin{figure*}
    \centering
    \includegraphics[trim=0 50 0 0,clip, width=0.85\linewidth]{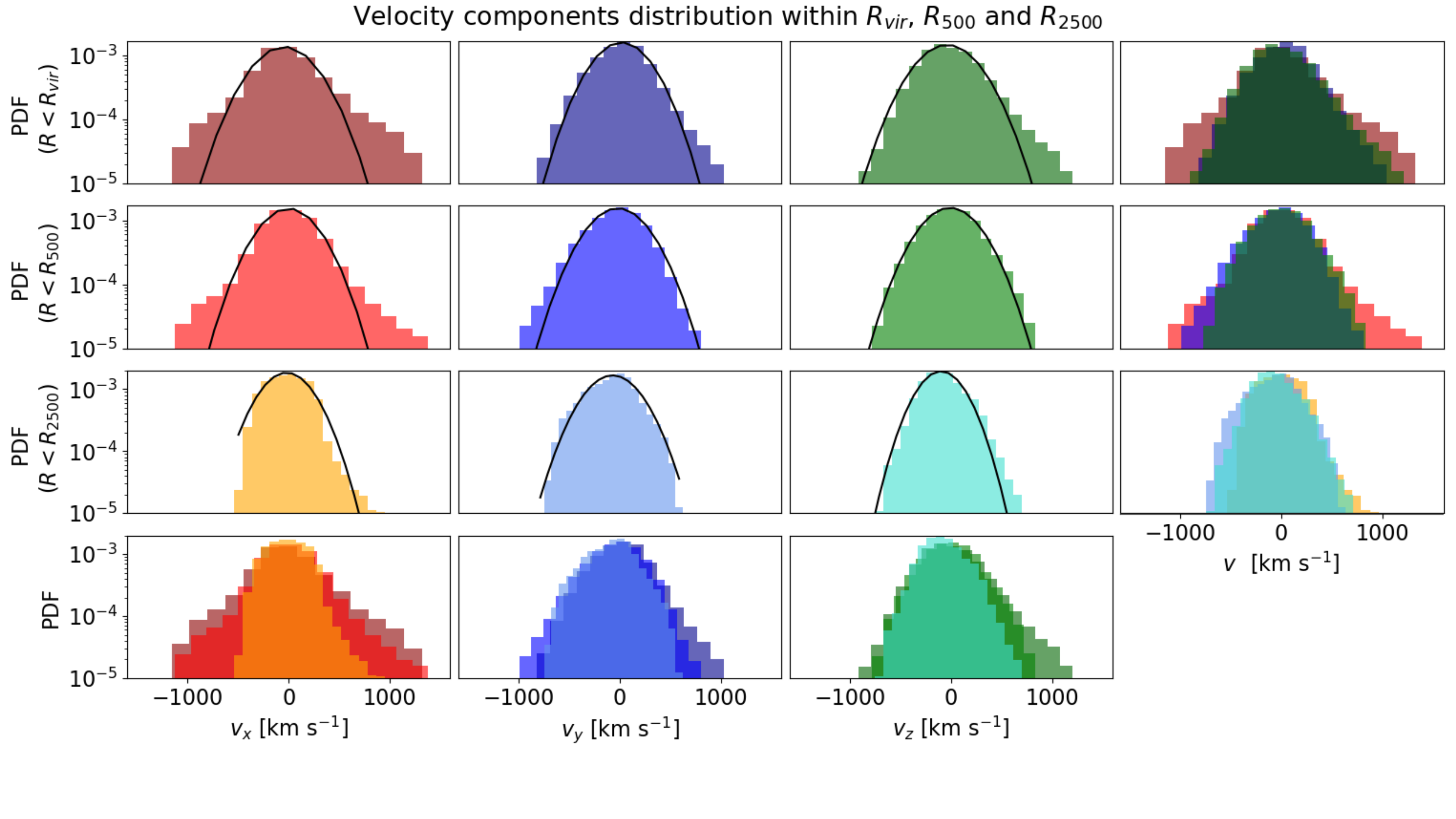}
    \caption{PDFs of the three components of the velocity field within $R_{vir}=2147$~kpc, $R_{500}=1087$~kpc, and $R_{2500}=515$~kpc. The first three rows display the PDFs computed within spheres of decreasing radius from top to bottom. The first three columns display the PDF of each component of the velocity field, $v_\mathrm{x}$, from red to yellow, $v_\mathrm{y}$ in shades of blue, and $v_\mathrm{z}$ in shades of green, from left to right. The right column overlays the PDFs of the three velocity-field components in a given sphere, and the bottom row overlays the PDFs of a given velocity-field component inside the different spheres. Fitted Gaussian envelopes are displayed as solid black curves. The statistical moments of the PDFs and the best-fit values of the Gaussian fit can be found in Table \ref{tab:stats PDFs}.}
    \label{pdf_3D}
\end{figure*}

\section{Statistical properties of the velocity field}
\label{sec:3 stat}

In this section, we study the statistical properties of the velocity field. We first analyse the probability distribution functions and their moments for both the 3D velocity field and the four projections presented above. To sample different configurations using a single simulated cluster, we also considered 100 randomly oriented projections.

As a first approximation, we expect that the dynamics in the ICM are only due to the orbital motion inside the gravitational potential well of the cluster. Thus, the PDFs of the velocity field are expected to follow a zero-centred Gaussian distribution in the cluster's rest frame. This should be the case for the 3D velocity field, but also for projected sightline velocities if there are no projection effects. We also expect isotropy; i.e. the three components of the velocity field should follow a similar distribution, including higher order statistics. Finally, if the velocity field is homogeneous, it should have a similar distribution, whatever the selected cluster region, provided the latter is large enough to be statistically representative. \\

\subsection{3D PDFs}

\begin{figure*}
    \centering
    \includegraphics[ width=0.85\linewidth]{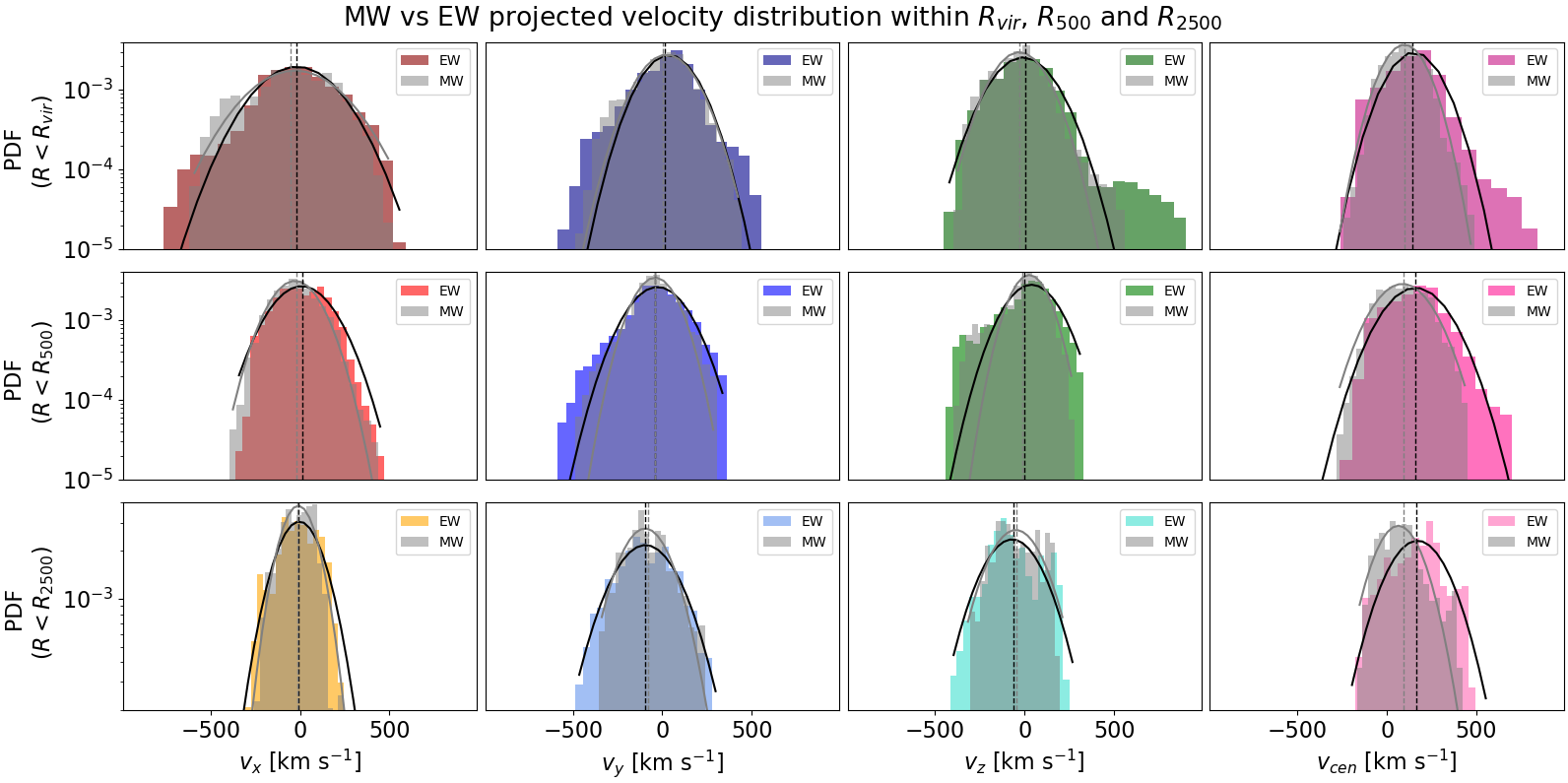}
    \caption{2D PDFs of EW and MW maps computed within circles of the radii $R_{vir}$, $R_{500,}$ and $R_{2500}$ from top to bottom, for the $v_\mathrm{x}$, $v_\mathrm{y}$, $v_\mathrm{z,}$ and $v_\mathrm{cen}$ projected velocities from left to right. The PDFs extracted from the EW maps are displayed following the same colour-code as Fig. \ref{pdf_3D}. The fitted Gaussian envelope is also shown as a solid black curve, and a vertical dashed black line represents the mean of the PDF. In addition, the PDFs extracted from the MW maps, their fitted Gaussian envelope, and their mean are overlaid in grey in the same style as their EW counterpart.}
    \label{pdf_2D}
\end{figure*}

We first studied the properties of the 3D velocity field by computing the PDFs of the three components of the velocity field within $R_\mathrm{vir}=2147$~kpc, $R_\mathrm{500}=1087$~kpc, and $R_\mathrm{2500}=515$~kpc. These are presented in Fig. \ref{pdf_3D}. Moreover, we calculated the following mass-weighted statistical properties: median, mean ($\mu$), standard deviation ($\sigma$), skewness, and kurtosis. They are reported in the top left of Table \ref{tab:stats PDFs}. A median different from the mean shows that the distribution is asymmetric, and a positive (negative) skewness indicates that there is more weight in the right (left) tail of the distribution, which means more positive (negative) velocities in our case. The kurtosis indicates the flatness or peakedness; a positive (negative) value indicates a flattened (peaked) distribution with more (less) weight in the tails compared to a Gaussian distribution. To compare with the statistical moments and visualise the distributions better, we fitted the PDFs to a Gaussian law, similarly to \cite{gatuzz2023measuring}. The best-fit values are reported on the top right of Table \ref{tab:stats PDFs}, and the Gaussian envelopes are displayed as solid black curves in Fig. \ref{pdf_3D}.\\

We first observe that, whatever the component and the sphere radius, the distributions are all roughly centred on zero. Although they are slightly shifted negatively, particularly for the $v_\mathrm{y}$ and $v_\mathrm{z}$ components within $R_\mathrm{2500}$. $\mu$ varies between -118.3 and 2.4~$\mathrm{km\,s^{-1}}$ most when the velocities are distributed in the [-1000,1000]~$\mathrm{km\,s^{-1}}$ range and $\sigma$ is scattered around 250~$\mathrm{km\,s^{-1}}$. Then, on the one hand, we note that the PDF of the $v_x$ component within $R_\mathrm{vir}$ and $R_\mathrm{500}$ has a lot of weight in the tails, as shown by its high kurtosis. It is due to two filaments, almost aligned with the $x$-axis of the simulation box (see Fig.1 of \citeauthor{2025A&A...704A..14Lebeau} \citeyear{2025A&A...704A..14Lebeau}), that funnel gas at high velocity towards the cluster. This gas being accreted is certainly not yet thermalised within Virgo's potential well. It also induces the highest velocity dispersion among the 3D PDFs: 325.6~$\mathrm{km\,s^{-1}}$. Indeed, the Gaussian fit yields a much smaller value of $\sigma=265.0\pm 11.1~\mathrm{km\,s^{-1}}$. On the other hand, the PDFs of the $v_\mathrm{y}$ and $v_\mathrm{z}$ components are tighter, but with a bit more weight in their right tail within $R_\mathrm{vir}$, as probed by their positive skewness. And, inversely, a bit more weight is in the left tail for the $v_\mathrm{y}$ component within $R_\mathrm{500}$. For these three PDFs, the fitted velocity dispersion is lower than that directly estimated from the PDFs, which is expected given their deviation from Gaussianity. Only the PDF of the $v_\mathrm{z}$ component within $R_\mathrm{500}$ shows an almost Gaussian distribution, with close-to-zero skewness and kurtosis and a fitted velocity dispersion in agreement with that of the PDF. \\

When comparing the PDFs of a component within different radii (bottom row of Fig.~\ref{pdf_3D}), we observe a narrowing of the distribution when reducing the radius of the sphere. In particular, within $R_\mathrm{2500}$, the tails are less populated. This may be because there is less non-thermal gas motion in the cluster core, even though this Virgo replica is known to be unrelaxed and hosting an active AGN in its core with a growing accretion shock (see \citeauthor{2024A&A...682A.157Lebeau} \citeyear{2024A&A...682A.157Lebeau}). This sphere might also be too small to characterise all the scales of the velocity field. We finally compare the PDFs of the different components within the same sphere. This confirms that, within $R_{vir}$, the velocity field is not isotropic because of the high-velocity gas accretion from the cluster's outskirts. On the contrary, the velocity field seems relatively isotropic within $R_\mathrm{2500}$, with quite similar PDFs for the three components. However, it might not be representative of the velocity field in the ICM. Thus, choosing to study the velocity field within $R_\mathrm{500}$ might be an acceptable compromise to recover representative statistical properties of the ICM without them being contaminated by large-scale accretion and biased by possible inhomogeneities or the lack of statistics within a smaller sphere. \\

\begin{figure*}[h!]
    \centering
    \includegraphics[trim=0 560 0 0,clip ,width=0.95\linewidth]{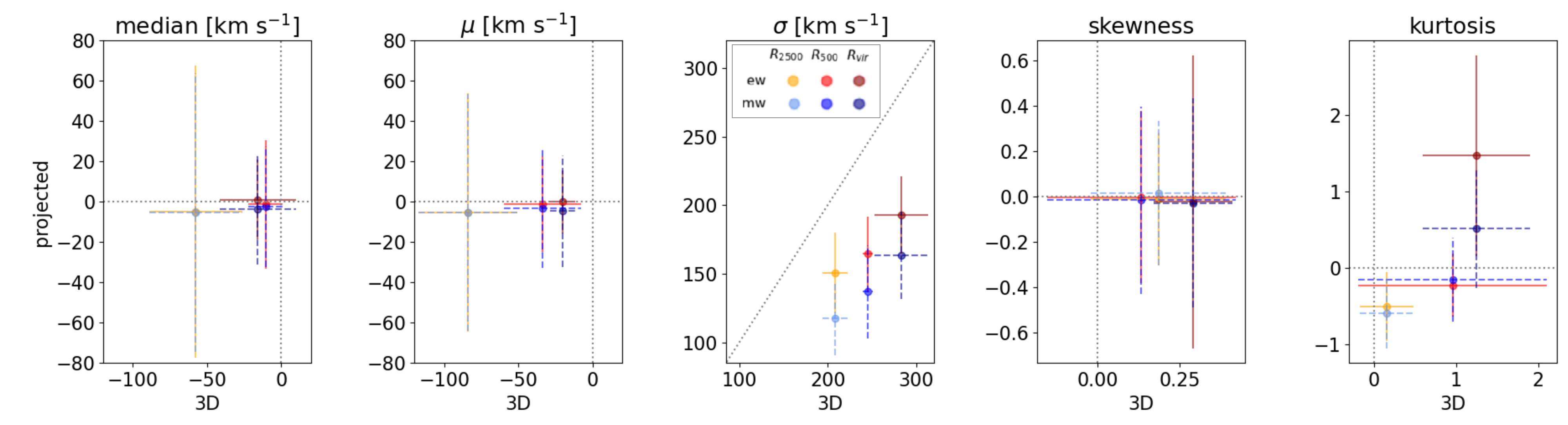}
    \caption{Comparison of statistical properties of the 3D velocity field (x-axis) over projected sightline velocities (y-axis). We show the mean value and the standard deviation among the three components of the 3D velocity field, i.e. $v_\mathrm{x}$,$v_\mathrm{y}$,$v_\mathrm{z}$, in the x-axis, and among 100 random projections in the y-axis. From left to right, we present the median, the mean, the standard deviation($\equiv\sigma_{proj}$ in the top panel of Fig. \ref{sigmas}), the skewness, and the kurtosis. We display the quantities estimated from EW (MW) projections within $R_\mathrm{2500}$ in orange (cyan), $R_\mathrm{500}$ in red (blue), and $R_\mathrm{vir}$ in dark red (dark blue). The legend can be found in the central panel.}
    \label{stats_random}
\end{figure*}

\subsection{2D PDFs of four projections}

We conducted the same analysis with the PDFs of the projected sightline velocity maps shown in Fig. \ref{los vel maps}. They are presented in Fig. \ref{pdf_2D}, similarly to the 3D velocity field (see Fig. \ref{pdf_3D}). The statistical moments and the best-fit values of these PDFs are reported in the middle (bottom) part of Table \ref{tab:stats PDFs} for the EW (MW) projections. \\

We first examine the mean and the median values; the PDFs of the EW and MW projections are close for each case, except for the $v_\mathrm{cen}$ projection and especially within $R_\mathrm{2500}$. However, they are, overall, more scattered than for the 3D PDFs. As shown by the visual inspection of Fig. \ref{los vel maps} conducted in Sect. \ref{sec:2 methodo}, the $v_\mathrm{x}$ projection is bimodal due to the gas accreted from two opposite filaments, leading to the largest dispersion among the projections. Then, the $v_\mathrm{y}$ and $v_\mathrm{z}$ projections are relatively zero-centred and have smaller dispersions. The $v_\mathrm{cen}$ projection is dominated by positive velocities due to the alignment with the filaments and the higher gas accretion from the background filament. The mean is about 150 (90)~$\mathrm{km\,s^{-1}}$ for the EW (MW) projections, whatever circle the PDFs are estimated in. \\

We also notice that, regardless of the projection, the weighting method, and the circle radius, the 2D PDFs are tighter than their 3D counterpart. Their standard deviation is roughly 100~$\mathrm{km\,s^{-1}}$ or more smaller for both the computed and the best-fit value. Moreover, we observe that the PDFs of the MW projections have much less weight in the tails than their EW counterparts, which is expected given that the MW method tends to smooth the projections when the EW method enhances the density peaks and thus the massive objects potentially falling faster onto Virgo than the overall gas flow. Consequently, the standard deviation is slightly smaller for the MW projections, except for the $v_\mathrm{x}$ projection within $R_\mathrm{vir}$. Accordingly, the kurtosis is also smaller for the MW projections for most cases. \\

There are also strong deviations from Gaussianity for the EW projection, in particular for the $v_\mathrm{x}$ and $v_\mathrm{z}$ projections within $R_\mathrm{vir}$, and the $v_\mathrm{cen}$ projection within $R_\mathrm{vir}$ and $R_\mathrm{500}$. For the $v_\mathrm{z}$ projection, it is because of a very high velocity and density region on the upper left of the map. It is certainly the gas accreted from the background filament reaching its highest velocity due to gravitational attraction just before being shocked when entering the cluster (as seen in Fig. 1 of \citeauthor{2025A&A...704A..14Lebeau} \citeyear{2025A&A...704A..14Lebeau}). For the $v_\mathrm{x}$ and $v_\mathrm{cen}$ projections, the important weight in the tails of their PDFs is also due to dense regions enhanced by the EW method, which are certainly halos falling onto Virgo. \\

In conclusion, there are important projection effects on the sightline velocity projections, particularly for the $v_\mathrm{x}$ and $v_\mathrm{cen}$ projections, because of filaments along the sightline. More importantly, they do not have the same statistical properties as their 3D counterparts, particularly the standard deviation.

\subsection{Generalisation to 100 randomly oriented projections}

\begin{figure*}[h]
    \centering
    \includegraphics[trim=0 0 0 0,clip,width=0.8\linewidth]{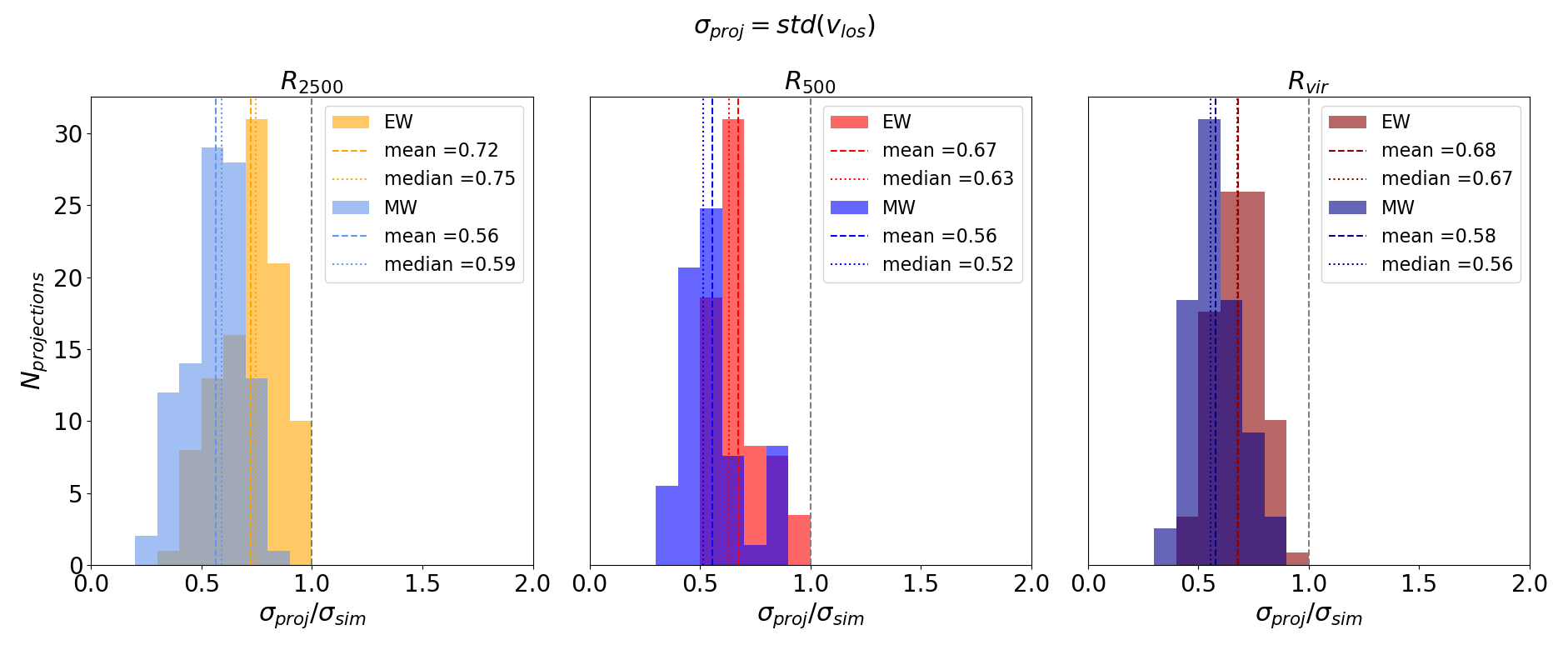}
    \includegraphics[trim=0 0 0 0,clip,width=0.8\linewidth]{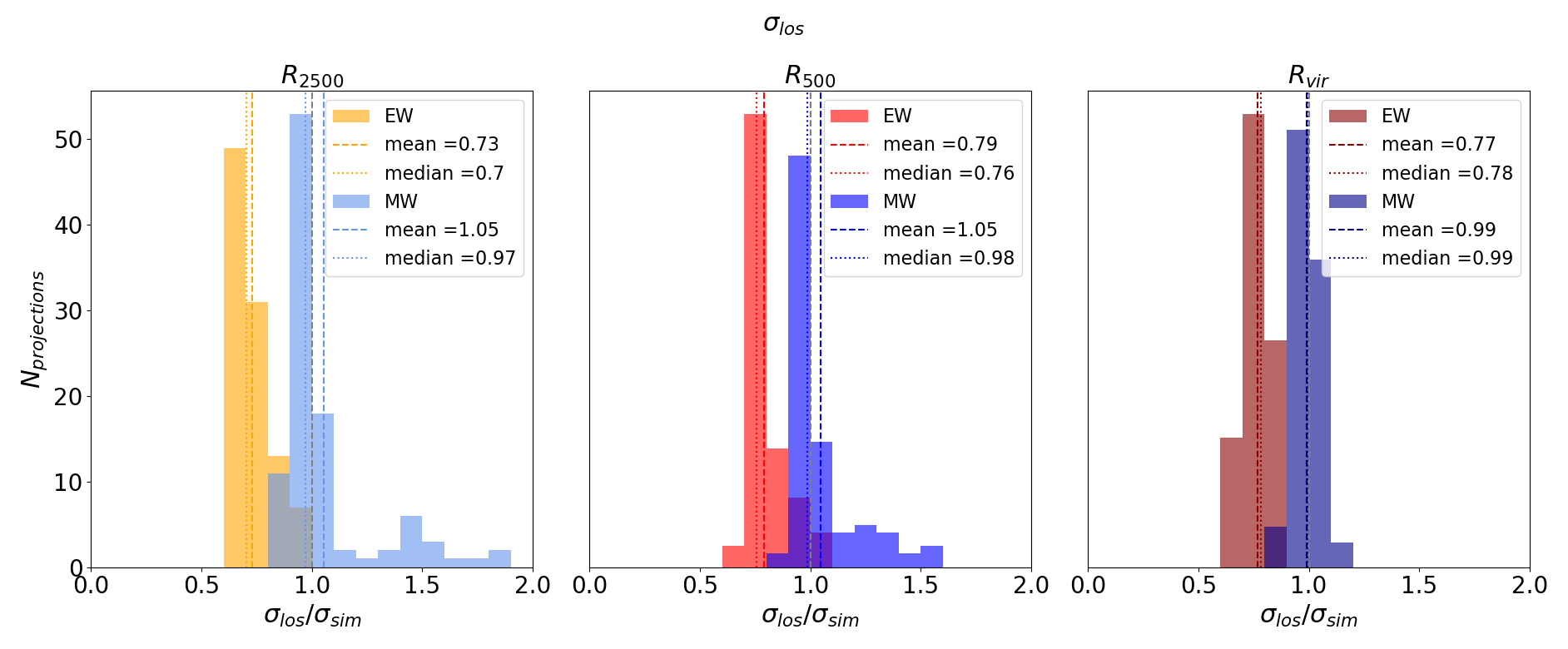}
    \caption{Histograms of the ratio of the standard deviationof sightline velocity ($\sigma_{proj}$, top, already presented in central panel of Fig. \ref{stats_random}, std stands for standard deviation) and sightline velocity dispersions ($\sigma_\mathrm{los}$, bottom) of one hundred random projections over the 1D velocity dispersion of the 3D velocity field, $\sigma_\mathrm{sim}^2=(\sigma_\mathrm{x}^2+\sigma_\mathrm{y}^2+\sigma_\mathrm{z}^2)/3$. For each row, we present the ratio of the quantities estimated within $R_\mathrm{2500}$ in orange for the EW projections (cyan for MW), $R_\mathrm{500}$ in red for the EW projections (blue for MW) and $R_\mathrm{vir}$ in dark red for the EW projections (dark blue for MW). The mean (median) of the distributions is displayed as dashed (dotted) vertical lines in the same colours and can be compared to one displayed as a dashed grey vertical line.}
    \label{sigmas}
\end{figure*}

We then generalised our method for 100 projections along random sightlines, as was done in \citet{2024A&A...689A..19Lebeau}. Figure \ref{stats_random} compares the mean statistical properties of the components of the 3D velocity field in the x-axis, and those of the 100 random projections are on the y-axis. From left to right, the median, mean, standard deviation, skewness and kurtosis are displayed. Along the x-axis and y-axis, the error bars are, respectively, the dispersion over the three components of the 3D velocity field and over the 100 projections. To test correlations between data points, we ran a bootstrap sampling. The histogram of the mean values of the sub-samples for each statistical moment and radius shows a central value in very good agreement with the values presented in Fig. \ref{stats_random}, and the dispersion is largely within the error bars presented in Fig. \ref{stats_random}.\\

We first notice that the median and mean of the projected velocities are scattered around zero; see the first two panels from the left. Their 3D counterparts within $R_\mathrm{500}$ and $R_\mathrm{vir}$ are close to 0~$\mathrm{km~s^{-1}}$, showing that it is almost at rest with respect to the halo velocity, which we consider as Virgo's rest frame. However, within $R_\mathrm{2500}$, the 3D velocity is around -50~$\mathrm{km~s^{-1}}$, certainly due to the gas flow triggered by the AGN feedback in Virgo's core, which is coherent with the large scatter among the projected velocity within this radius. The skewness of the projected velocities is also centred on zero, whatever the radius, with the highest scatter within $R_\mathrm{vir}$ being due to the matter accretion from filaments. For the 3D velocity field, it is slightly positive, whatever the radius, meaning that the distribution is slightly asymmetric towards higher velocities. The kurtosis indicates that the larger the radius, the flatter the distribution for both the 3D and the projected velocities. It is due to the accretion at high velocities in the outskirts and the low velocities in the core.  \\

Overall, for these four quantities, the weighting method does not have much impact. On the contrary, we observe a clear trend for the standard deviation: it is higher for EW projections than for MW ones and decreases almost linearly with decreasing radius. More importantly, as written above, it is always smaller than the standard deviation of the ground-truth 3D-simulated velocity field. The one-to-one relation is shown as a dotted diagonal grey line. To investigate this observation in more detail, we computed the ratio of the standard deviation of the sightline velocity projections, named $\sigma_\mathrm{proj}$, over the 1D velocity dispersion of the 3D-simulated velocity field, which is the root mean square (rms) of the dispersion of the components: $\sigma_\mathrm{sim}=\sqrt{(\sigma_\mathrm{x}^2+\sigma_\mathrm{y}^2+\sigma_\mathrm{z}^2)/3}$. The histogram of this ratio over the 100 random projections is presented in the top panel of Fig. \ref{sigmas}. The panels present the distributions within $R_\mathrm{2500}$, $R_\mathrm{500,}$ and $R_\mathrm{vir}$ from left to right. The distribution of the ratios, and thus the underlying standard deviation, looks relatively Gaussian within the three radii since their mean and median agree well. The histograms confirm that the one hundred sightline velocity projections, both EW and MW, have a smaller velocity dispersion than the ground-truth simulated velocity field, by a factor of around 0.7 for the EW projections and 0.55 for the MW projections, regardless of the radius. \\

We can also access the sightline velocity dispersion from the broadening of spectral lines in the X-rays, which we modelled following \citet{roncarelli2018athena} (see Fig. \ref{los vel disp maps}). Following the same generalisation strategy, we produced 100 MW and EW sightline velocity-dispersion projections using the same random angles as the sightline velocity projections. The bottom row of Fig. \ref{sigmas} presents histograms of the ratio of the mean sightline velocity dispersion, $\sigma_\mathrm{los}$ (using Eq. \ref{equ:sigma}), over the mean 3D velocity dispersion in the same way as in the top row.  We can observe that, whatever the radius, almost all the sightline velocity-dispersion ratios, and hence the mean and median, are very close to one for MW projections. In contrast, they are biased and low for EW projections, with a mean and median around 0.75. This result is in very good agreement with that of \citeauthor{2026A&A...705A.129Vazza} (\citeyear{2026A&A...705A.129Vazza}, see top panel of their Fig.4), which found that the X-ray-weighted (labelled EW in our analysis) sightline velocity dispersion is lower than the volume-weighted (MW) value, and it could explain the quite low sightline velocity dispersion observed by XRISM. Consequently, we can argue that in the case of Virgo the MW sightline velocity dispersion seems to reproduce the velocity dispersion of the ground-truth 3D simulated velocity field the most accurately. We can also argue that the standard deviation (std) of the sightline velocity projections, $\sigma_\mathrm{proj}=std(v_\mathrm{los})$, is not equivalent to the sightline velocity dispersion projections, $\sigma_\mathrm{los}$, which is supposed to be in the ideal case of isotropy and unbiased projections. We compare these results to other works in Sect. \ref{discussion}.

\section{Turbulent Mach number and non-thermal pressure}
\label{sec:4 Mach}

\begin{figure*}
    \centering
    \includegraphics[trim= 0 280 350 120,clip, width=0.85\linewidth]{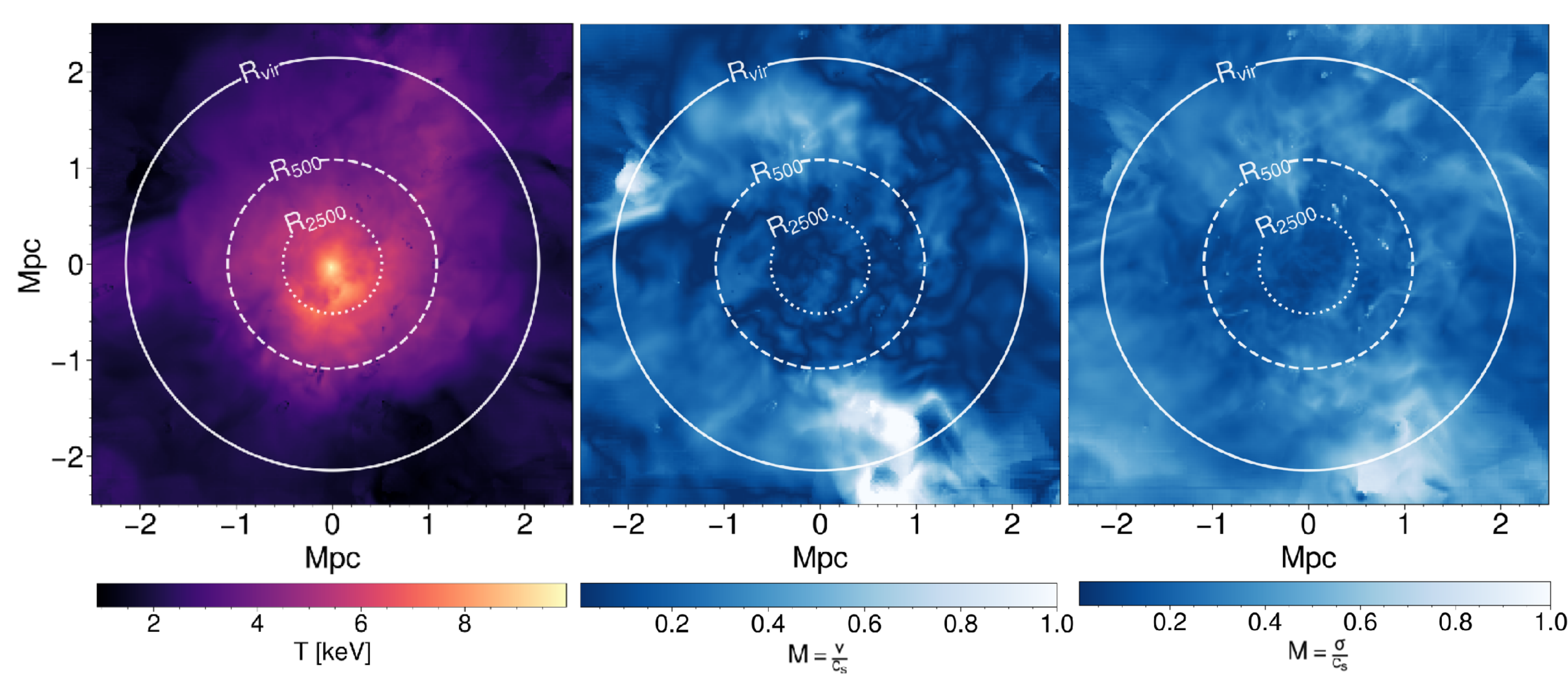}
    \caption{Projected spectroscopic-like temperature (left) and the ratio of the EW sightline velocity (centre) and velocity dispersion (right) over the sound speed in each pixel, which is thus a local Mach number, along the $x$ sightline. The solid, dashed and dotted white circles respectively represent $R_\mathrm{vir}$, $R_\mathrm{500}$ and $R_\mathrm{2500}$.}
    \label{t mach maps}
\end{figure*}

After inspecting the properties of the velocity field, we then derived the level of non-thermal pressure due to the gas motion using the effective turbulent Mach number, as proposed in \citet{eckert2019non} and recently used in \citet{2025ApJ...982L...5XRISM,2025PASJ...77S.242XRISM,2025ApJ...985L..20XRISM-Coma}. From the estimates of the velocity field presented in the previous sections, we derived the level of non-thermal pressure due to the gas motions: 

\begin{equation}
   \alpha=\frac{P_\mathrm{nt}}{P_\mathrm{tot}}=\frac{P_\mathrm{nt}}{P_\mathrm{th}+P_\mathrm{nt}}=\frac{\frac{1}{3}\rho\sigma_\mathrm{3D}^2}{\frac{\rho k_BT}{\mu m_p}+\frac{1}{3}\rho\sigma_\mathrm{3D}^2}=\frac{M_\mathrm{3D}^2}{\frac{3}{\gamma} +M_\mathrm{3D}^2}
   \label{equ:alpha}
,\end{equation}where the non-thermal pressure is defined as $P_\mathrm{nt}=\frac{1}{3}\rho_\mathrm{gas}\sigma^2_\mathrm{3D}$, with $\rho_\mathrm{gas}$ being the gas density and $\sigma^2_\mathrm{3D}=\sigma_\mathrm{x}^2+\sigma_\mathrm{y}^2+\sigma_\mathrm{z}^2$ the 3D velocity dispersion \citep[as in e.g.][]{2012MNRAS.422.2712Zhuravleva,nelson2014hydrodynamic,eckert2019non,angelinelli2020turbulent}. Since isotropy is assumed, we also have $\sigma_\mathrm{3D}^2=\sigma_\mathrm{1D}^2/3$ (see e.g. \citeauthor{biffi2016nature} \citeyear{biffi2016nature}), with $\sigma_\mathrm{1D}$ being equal to $\sigma_\mathrm{sim}$ for the 3D-simulation case and equal to $\sigma_\mathrm{los}$ and $\sigma_\mathrm{proj}$ for the projected approaches. \\

In Eq. \ref{equ:alpha}, we assume that velocity dispersion dominates over the velocity norm in the effective turbulent Mach number $M_\mathrm{3D}=(\sigma^2_\mathrm{3D}+\overline{v}^2)^{1/2}/c_s\approx\sigma_\mathrm{3D}/c_s$, which is a good approximation in our case (see Table \ref{app: pnt_mach tab}). It includes $c_s=(\gamma k_BT/\mu m_p)^{1/2}$, which is the sound speed, with $\gamma=5/3$ being the adiabatic index, $\mu=0.6$ the mean molecular weight, and $m_p$ the proton mass. This approach has the advantage of only needing the temperature, sightline velocity, and sightline velocity dispersion to estimate the $\alpha$ parameter, which are all accessible in X-ray observations. However, in observations we do not have access to the 3D velocity field and its dispersion; we only have the sightline velocity, in the cluster's rest frame, using the BCG's redshift as reference, and the velocity dispersion. By assuming isotropy, we thus have $M_\mathrm{3D}=(3\sigma^2_\mathrm{1D}+\overline{v}^2)^{1/2}/c_s$, with $\overline{v}^2$ being the sightline velocity or the velocity norm for the 3D case, and $\sigma_\mathrm{1D}$ as described above. \\

We estimated these quantities from 3D simulation outputs by computing their mass-weighted mean within $R_\mathrm{500}$ and $R_\mathrm{vir}$. Then, in the observational approach, the projected sightline velocity and velocity dispersion were already presented above, and the projected temperature was computed using the spectroscopic-like weighting of \citet{2004MNRAS.354...10Mazzotta}:

\begin{equation}
    T_{\mathrm{sl}}=\frac{\sum_\mathrm{i} T_\mathrm{i}w_\mathrm{i}}{\sum_\mathrm{i}w_\mathrm{i}}, \quad w_\mathrm{i}=\frac{n_{e_\mathrm{i}}}{T_\mathrm{i}^{3/4}}
.\end{equation}We estimated the mean of these projected quantities within $R_\mathrm{500}$ and $R_\mathrm{vir}$ for the $x$,$y$,$z$ and $cen$ projections. Figure \ref{t mach maps} presents, from left to right, the temperature and the ratio of the EW sightline velocity and velocity dispersion over the sound speed in each pixel, which is thus a local Mach number, along the $x$ sightline as an example. We can see that the temperature rises up to more than 8~keV within $R_\mathrm{2500}$, is between 4 and 8~keV in the $[R_\mathrm{2500}, R_\mathrm{500}]$ range, and is below 4~keV beyond. However, it is important to note that this simulation of Virgo experienced a major merger about 0.5~Gyr before z=0; thus, the core of Virgo is not relaxed yet. The local Mach numbers show that, as already discussed, the velocity dispersion dominates the energy content in the core as its associated Mach number is in the [0.4,0.8] range within $R_\mathrm{500}$ when that of the sightline velocity is in the [0,0.4] range. On the contrary, beyond $R_\mathrm{500}$, the Mach number of the sightline velocity rises to one due to the accretion of matter from filaments, while that of velocity dispersion rises to 0.8. In addition, the Mach number highlights the characteristic scales of velocity fluctuations in correlation with the temperature fluctuations in the cluster core, as we see patterns of similar sizes, particularly for the sightline velocity (central panel).\\

\begin{figure*}
    \centering
    \includegraphics[trim=0 20 0 350,clip,width=1\linewidth]{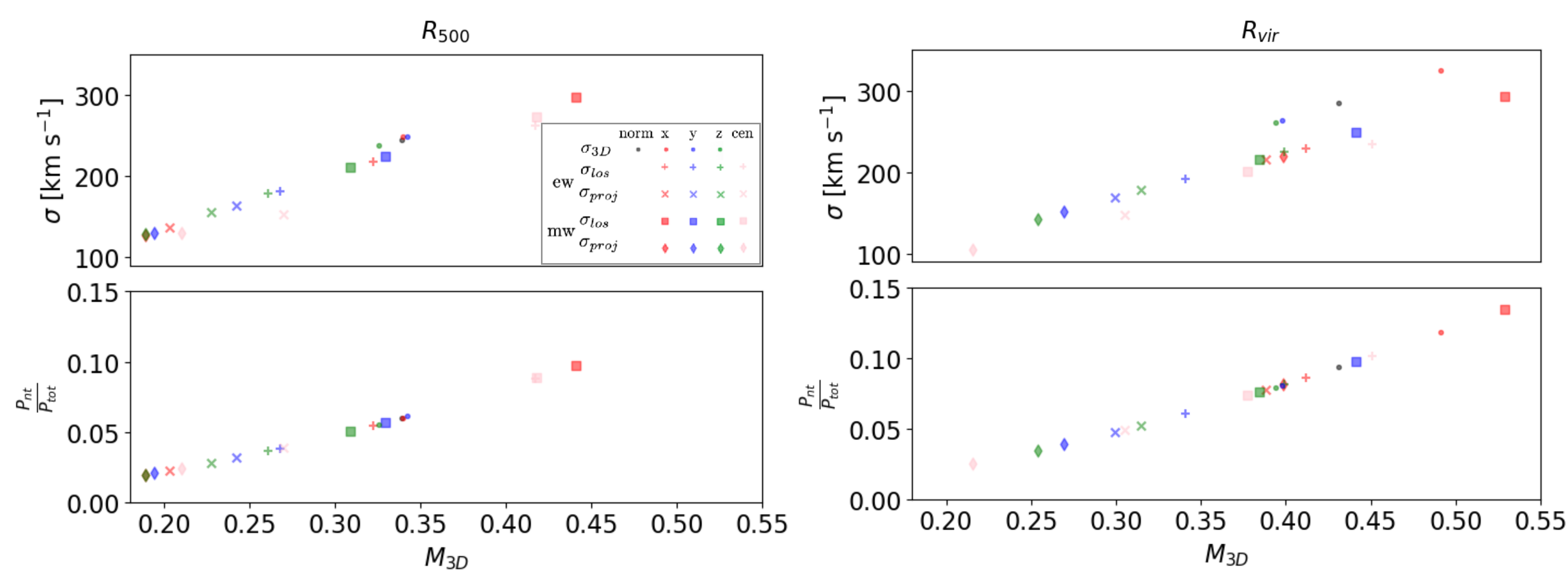}
    \caption{Mach number (x-axis) compared to the velocity dispersion, $\sigma$, (top) and the non-thermal pressure fraction (bottom). The left (right) panel presents the quantities computed within $R_\mathrm{500}$ ($R_\mathrm{vir}$). Each marker represents a method to compute the velocity dispersion: dots for the 3D, 'plus' (square) for the sightline velocity dispersion and cross (diamond) for the standard deviation over the sightline velocity, both from EW (MW) projections. The values derived from the norm of the 3D velocity field are displayed in black. The quantities derived from the $x$, $y$, $z$, and $cen$ components (indeed only $x$,$y$,$z$) or projections are respectively displayed in red, blue, green and pink.}
    \label{p_nt_mach}
\end{figure*}

We used both the MW and EW projections for the sightline velocity and velocity dispersion, $\sigma_\mathrm{los}$. In addition, we computed the standard deviation of the sightline velocity, $\sigma_{proj}$, as already presented in Fig. \ref{sigmas}. We present the results in Fig. \ref{p_nt_mach}. The left (right) panel presents the values estimated within $R_\mathrm{500}$ ($R_\mathrm{vir}$). The Mach number (x-axis) is compared to the velocity dispersion, $\sigma$ (top), and the non-thermal pressure fraction (bottom). All the values displayed in this figure can be found in Table \ref{app: pnt_mach tab}. \\

For both radii, the temperature estimated from the 3D simulation outputs is higher than that calculated from spectroscopic-like projected temperatures (see Table \ref{app: pnt_mach tab}). This leads to a sound speed higher than 1300 (1200)~$\mathrm{km~s^{-1}}$ within $R_\mathrm{500}$ ($R_\mathrm{vir}$), whereas it is below 1200 (1000)~$\mathrm{km~s^{-1}}$ for the projected counterparts. Nevertheless, this difference only leads to a difference of approximately a few percent in the Mach number estimation for a fixed velocity dispersion of, for example, 250~$\mathrm{km~s^{-1}}$. Thus, the velocities are distributed around 0~$\mathrm{km~s^{-1}}$ (see also Table \ref{app: pnt_mach tab}), except for the $cen$ projection, which is around 100~$\mathrm{km~s^{-1}}$, for both radii, due to matter accretion from the aligned background filament. Whatever the value of the velocity, it is always much lower than the velocity dispersion, which is distributed roughly between 100 and 300~$\mathrm{km~s^{-1}}$ (see top panels of Fig. \ref{p_nt_mach}) for both radii and thus dominates over the velocity. We clearly see this by the almost linear correlation between $\sigma$ and $M_\mathrm{3D}$. There is also an expected correlation between $P_\mathrm{nt}/P_\mathrm{tot}$ and $M_\mathrm{3D}$ on the bottom sub-panels of Fig. \ref{p_nt_mach}. For the 3D components and norm, we have a Mach number of about [0.32,0.35] ([0.4,0.5]) within $R_\mathrm{500}$ ($R_\mathrm{vir}$), leading to a non-thermal pressure fraction of around $6\%$ ($9\%$). For the projected values, on the one hand, in both radii and for both weighting methods, the sightline velocity dispersion of $\sigma_\mathrm{los}$ is (as expected) in good agreement with the 1D velocity dispersion of the 3D velocity field, and so are the Mach number and the non-thermal pressure fraction. On the other hand, $\sigma_\mathrm{proj}$ is much lower (as shown in the top panel of Fig. \ref{sigmas}), so the estimated Mach number and non-thermal pressure fraction are also much lower in the 2.5 to 4$\%$ range (3 to 6 $\%$) within $R_\mathrm{500}$ ($R_\mathrm{vir}$). The three exceptions are the $x$ 3D component and the $x$ and $cen$ projections, for which we show significant projection effects due to accretion from filaments; these lead to a very large velocity dispersion, and hence to a non-thermal pressure fraction up to 10$\%$ within $R_\mathrm{500}$. Within $R_\mathrm{vir}$, only the $x$ component and projection have an overestimated velocity dispersion of about 300~$\mathrm{km~s^{-1}}$, leading to a non-thermal pressure fraction of 15$\%$. \\

The Mach numbers estimated within $R_{500}$, either from 3D or sightline velocity dispersion, are in rather good agreement with the values estimated by \citet{2022MNRAS.511.4511GatuzzVirgo}, though the latter are estimated within a very central region, as we discuss further in Sect. \ref{discussion}. Moreover, the fraction of non-thermal pressure, even for our $x$ component and projection, is in good agreement with most of the values found by \citet{eckert2019non} using the X-COP sample at $R_\mathrm{500}$. The exception is A2319, which has a non-thermal pressure fraction higher than 40$\%$ at both radii. Similarly to \citet{eckert2019non}, our values are slightly below those of \citet{nelson2014hydrodynamic} and The300, but also \citet{gianfagna2021exploring} and those of \citet{2025ApJ...984L..63Sarkar} obtained with a joint SPT\footnote{\url{https://pole.uchicago.edu/public/Home.html}} and XMM-Newton\footnote{\url{https://www.cosmos.esa.int/web/xmm-newton}} analysis. 

\section{Velocity structure function}
\label{sec:VSF}

After estimating the non-thermal pressure fraction, which is assumed to result mostly from turbulence, we now use the velocity structure function (VSF) to characterise the structuration of the velocity distribution across scales and thus assess the presence of the turbulent cascade. The n-th order VSF, written $\delta v^n(\mathbf{r})$ hereafter, is the n-th moment of the absolute velocity difference of all data points separated by a distance, $\mathbf{r}$. In the core of this study, we only considered a constant weight (see Fig. \ref{app: VSF weighting compar} for a density-weighted VSF). The VSF is thus written as

\begin{equation}
    \delta v^n(\mathbf{r}) = \frac{\sum_\mathrm{i=1}^N |v(\mathbf{x_i}+\mathbf{r})-v(\mathbf{x_i})|^\mathbf{n}}{N}=<|v(\mathbf{x_i}+\mathbf{r})-v(\mathbf{x_i})|^\mathbf{n}>
,\end{equation}

\noindent with $v$ being the sightline velocity and N the number of data points separated by a distance, $\mathbf{r}$. \\

\begin{figure*}
    \centering
    \includegraphics[width=0.8\linewidth]{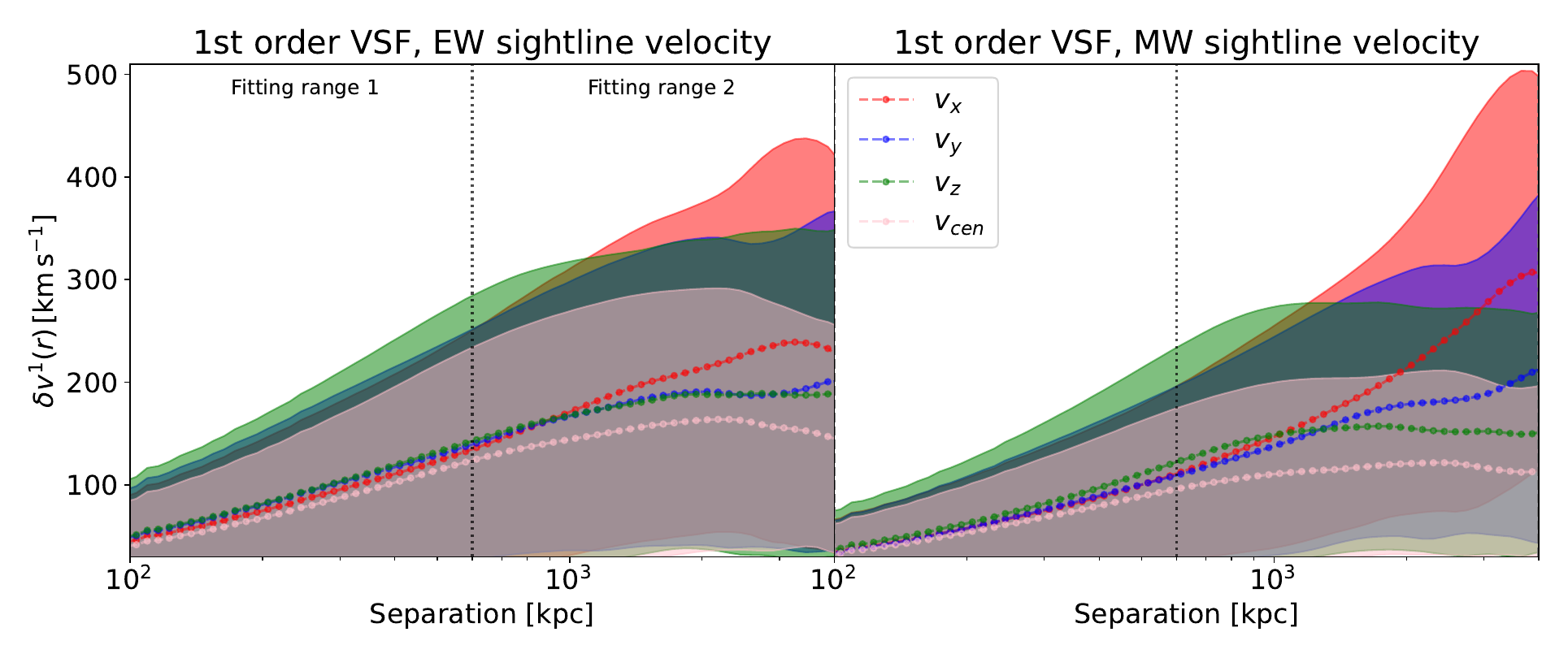}
    \caption{First order velocity structure function (VSF) computed from EW (left) and MW projected maps (right). They are displayed in red for the $v_\mathrm{x}$ projection, blue for the $v_\mathrm{y}$ projection, green for the $v_\mathrm{z}$ projection and pink for the $v_\mathrm{cen}$ projection. The dispersions are displayed in shaded areas in the same colours. They are presented in the [100,4000]~kpc range, considered as the range of validity of the study (see left panel of Fig. \ref{app: val range fig + vsf comp 222 444}. The range is separated into two fitting ranges: [100,600]~kpc and [600,1000]~kpc, separated by a dashed vertical dotted black line on the plot. The best-fit slope values in each interval are reported in Table \ref{tab:fit}.} 
    \label{VSF1}
\end{figure*}

We computed the first- and second-order VSFs. The first order is simply the mean absolute velocity difference and thus informs us of the general gas flow, while the second-order VSF is equal to twice the variance minus the correlation function (see e.g. \citeauthor{schulz1981structure} \citeyear{schulz1981structure}). The first- and second-order VSFs are respectively presented in Figs. \ref{VSF1} and \ref{VSF2}. For both figures, the VSF was computed from EW and MW maps, which are presented in the left panel for the former and the right panel for the latter. The comparison of the VSF computed from the EW and MW maps for a given projection can be found in Appendix \ref{app:vsf add figs}. The separation between pixels in the maps ranges between 22.5~kpc, which is a pixel width, and 7065~kpc, in the diagonal of the map. The VSFs were thus binned in 100 intervals on a log scale in this range. Given the small number of values in the bins at the shortest and longest separations (see Fig. \ref{app: val range fig + vsf comp 222 444}), we defined the range of validity between 100 and 4000~kpc for this analysis. \\

We observe that, regardless of the order and for both the EW and MW maps, the VSFs seem to have roughly the same slope in the [100,600]~kpc range, but behave very differently in the [600,4000]~kpc range. We named these two intervals 'Fitting range 1' and 'Fitting range 2', in which we fitted a power-law $\delta v^n(r)=Ar^{\alpha}$. A dotted vertical black line highlights the delimitation. The best-fit values of the slope, $\alpha,$ are reported in Table \ref{tab:fit}. It is worth noting that the uncertainties on the best-fit value of the slope are of the order of $10^{-4}$ and are thus not reported in the table. It confirms that the VSFs have very similar slopes in the first fitting range, although with a bit more dispersion for the second-order VSF of the MW projections.\\

In the [600,4000]~kpc range, the differences in slope reflect the contribution of the filaments connected to Virgo. As a matter of fact, the bimodality observed in the $v_\mathrm{x}$ projection induces large velocity differences at large separations, leading to the steepest VSF slope among the projections. We notice that the slope is steeper for the MW method because this method smooths the projection compared to the EW method, which has the combined effect of lowering the contribution of density peaks while leveraging the contribution of underdense regions. It is particularly visible at the bottom of the projection (left panels of Fig. \ref{los vel maps}), where the EW projection has a very negative velocity region, but the rest is closer to zero. In contrast, the MW projection has a velocity distribution of around $-400~\mathrm{km\,s^{-1}}$ in the same bottom part. It is also visible on the PDFs (top left panel of Fig. \ref{pdf_2D}). The EW projection has pixels with lower velocities than the MW projection, but has much less weight in the left tail. In contrast, the $v_\mathrm{cen}$ projection has the lowest slope. The VSF almost flattens in this range, which is expected because it is dominated by the gas flow from the background filament; this can be considered as coherent and thus has low-velocity gradients. The MW method has a lower slope for that projection than its EW counterpart, given that it just smooths the velocity field, which is not as contrasted as the $v_\mathrm{x}$ projection. Finally, the $v_\mathrm{y}$ and $v_\mathrm{z}$ projections have intermediate VSF slopes, although they tend to flatten beyond a 1~Mpc separation, which could mean that there is no turbulent cascade on these scales. As shown by the visual inspection of the projections in Fig. \ref{los vel maps} and the analysis of their PDFs in Fig. \ref{pdf_2D}, these two projections are the least impacted by the filaments connected to Virgo. They would thus be the most representative of an ideal, relaxed, and isolated cluster. \\

Given the similarity of the slope in all the projections in the first fitting range, we could consider it the inertial range where the turbulent cascade occurs. In that scenario, the injection scale, i.e. the size of the largest vortices before collapsing into smaller ones, would be in the [600,800]~kpc range. This typical scale is roughly the diameter of the AGN shockfront observed in \citet{2024A&A...682A.157Lebeau}, which could contribute to generating vortices and, thus, turbulence in Virgo's core. On the other hand, we cannot observe the dissipation scale at small separations, which we might observe on much better resolved projected maps, but it is beyond the scope of this study. The second fitting range highlights the level of contribution of local large-scale structures along some sightlines. In terms of slope, the Kolmogorov theory \citep{kolmogorov1941local} predicts $\delta v^n(r)\propto r^{n/3}$ in the inertial range for the n-th order VSF, whereas the computed VSFs are much steeper and seem closer to $\delta v^n(r)\propto r^{(n+1)/3}$. It might mostly be due to projection effects steepening the VSF, as observed by \citet{mohapatra2022characterizing} or \citet{2025A&A...698A.121Fournier}, for example, since we showed in \citet{2025A&A...704A..14Lebeau} that the 3D velocity power spectrum (which is the Fourier transform of the second-order VSF) in the core of the Virgo replica is in excellent agreement with the Kolmogorov prediction. This finding is in agreement with \citet{2026A&A...705A.129Vazza}, which showed that the 3D second-order VSF of their Coma-like simulated cluster is also in good agreement with Kolmogorov-like turbulence on a rather large range of scales.

\begin{table}
    \centering
    \caption{Best-fit values of the velocity structure function slope.}
    \renewcommand{\arraystretch}{1.1}
    \begin{adjustbox}{width=0.45\textwidth}
    \begin{tabular}{ c c c c c c }
    \multicolumn{6}{c}{Fitting of the VSF slope}\\
    \hline \hline
    \multicolumn{3}{c}{Kolmogorov : $\delta v^n(r)\propto r^n/3$} & \multicolumn{3}{c}{Fit : $\delta v^n(r)\propto r^{\alpha}$ } \\
    \hline \hline
    \multicolumn{6}{c}{Fitting range 1 : [100,600]~kpc}\\
    \hline
     VSF order (n) & weighting & $v_\mathrm{x}$ & $v_\mathrm{y}$ & $v_\mathrm{z}$ & $v_\mathrm{cen}$\\
    \hdashline 
     \multirow{2}{*}{1} & EW & 0.57 & 0.54 & 0.55 & 0.59 \\
     & MW &  0.61 & 0.62 & 0.63 & 0.57\\
    \hdashline 
    \multirow{2}{*}{2} & EW & 1.03 & 0.97 & 1.01 & 1.06 \\
    & MW & 1.32 & 1.12 & 1.19 & 1.06 \\
    \hline
    \multicolumn{6}{c}{Fitting range 2 : [600,4000]~kpc}\\
    \hline
     VSF order (n) & weighting & $v_\mathrm{x}$ & $v_\mathrm{y}$ & $v_\mathrm{z}$ & $v_\mathrm{cen}$\\
    \hdashline 
     \multirow{2}{*}{1} & EW & 0.29 & 0.15 & 0.13 & 0.09 \\
     & MW & 0.56 & 0.31 & 0.07 & 0.08 \\
    \hdashline 
    \multirow{2}{*}{2} & EW & 0.56 & 0.30 & 0.19 & 0.11 \\
    & MW & 1.09 & 0.58 & 0.07 & 0.09 \\
    
    \end{tabular}
    \end{adjustbox}
    \label{tab:fit}
\end{table}

\section{Discussion}
\label{discussion}

Throughout this work, we used both the EW and MW methods to project and quantify the velocity field of Virgo's ICM through different tracers. We explain the motivations for the use of these weighting methods and summarise the different results yielded by each of them.\\

On the one hand, the MW method is physically motivated when averaging over intensive physical quantities such as the pressure, temperature, or velocity, given that the cells or particles extracted from cosmological simulations have different masses. For instance, this method is widely used in analyses of direct outputs of simulations to compute radial profiles of several properties of galaxy clusters. We used it to project the pressure along multiple sightlines in \citet{2024A&A...682A.157Lebeau}; we thus logically tested this method on the velocity field in this work. On the other hand, the EW method is used to mimic observed quantities without fully reproducing them since the weight is the squared electron density (similarly to the X-ray's surface brightness); it is thus used in mock X-ray observations such as \citet{2012MNRAS.420.3545Biffi} or \citet{roncarelli2018athena}, including the sightline velocity and velocity dispersion as in this study. It was also shown by \citet{2004MNRAS.354...10Mazzotta} and \citet{2005ApJ...618L...1Rasia} that the spectroscopic-like weighting method is a better approximation than the EW one to reproduce the observed spectroscopic temperature. We thus followed this approach for the temperature. For the sightline velocity and velocity dispersion, there is not, to our knowledge, a thorough comparison between the MW and EW methods in the literature. \\

For the sightline velocity, we observed that the EW projections are more contrasted than their MW counterpart (see Fig. \ref{los vel maps}), thus leading to a larger standard deviation. However, in both cases, the standard deviation is much lower than its simulated counterpart, as shown in the top panel of Fig. \ref{sigmas}. In addition, the mean velocity extracted within $R_\mathrm{2500}$, $R_\mathrm{500,}$ or $R_\mathrm{vir}$ on these sightline velocities is centred on zero over the 100 random projections for both methods. The VSFs computed from the EW projections have a higher amplitude and slightly higher slope than the MW ones in the [100,600]~kpc range, as shown in Table \ref{tab:fit} and compared in Fig. \ref{app: VSF EW MW comp}, but the general trend remains the same. However, beyond 600~kpc, in the case of large projection effects induced by filaments such as in the $x$ projection, the smoother MW projection leads to a steeper VSF because the number of pairs with important velocity difference is larger than its more peaked EW counterpart. \\

For the sightline velocity dispersion (see Fig. \ref{los vel disp maps}), the EW projections are also more peaked, but the mean velocity dispersion is on average smaller than for MW projections (see bottom panel of Fig. \ref{sigmas}). It is inverted compared to the standard deviation of sightline velocity. This is due to the fact that EW sightline velocity is used to compute the EW sightline velocity dispersion (see second member of Eq. \ref{equ:sigma}). Since it is more contrasted, it yields a higher squared sightline velocity than its MW equivalent, leading to smaller sightline velocity dispersion than with the MW method. We observe that the MW $\sigma_\mathrm{los}/\sigma_\mathrm{sim}$ ratio is the closest to one at each radius, leading to the most accurate non-thermal pressure fraction estimation when considering the MW, direct simulation outputs as the ground-truth. Thus, in a simulation-oriented approach, like this work, the MW projections should be preferred to the EW ones to study the velocity field along sightlines, particularly when studying non-thermal pressure. However, it does not show that the MW method should be used instead of the EW method for mock observations. \\

\begin{figure}
    \centering
    \includegraphics[trim= 30 20 30 0,width=0.9\linewidth]{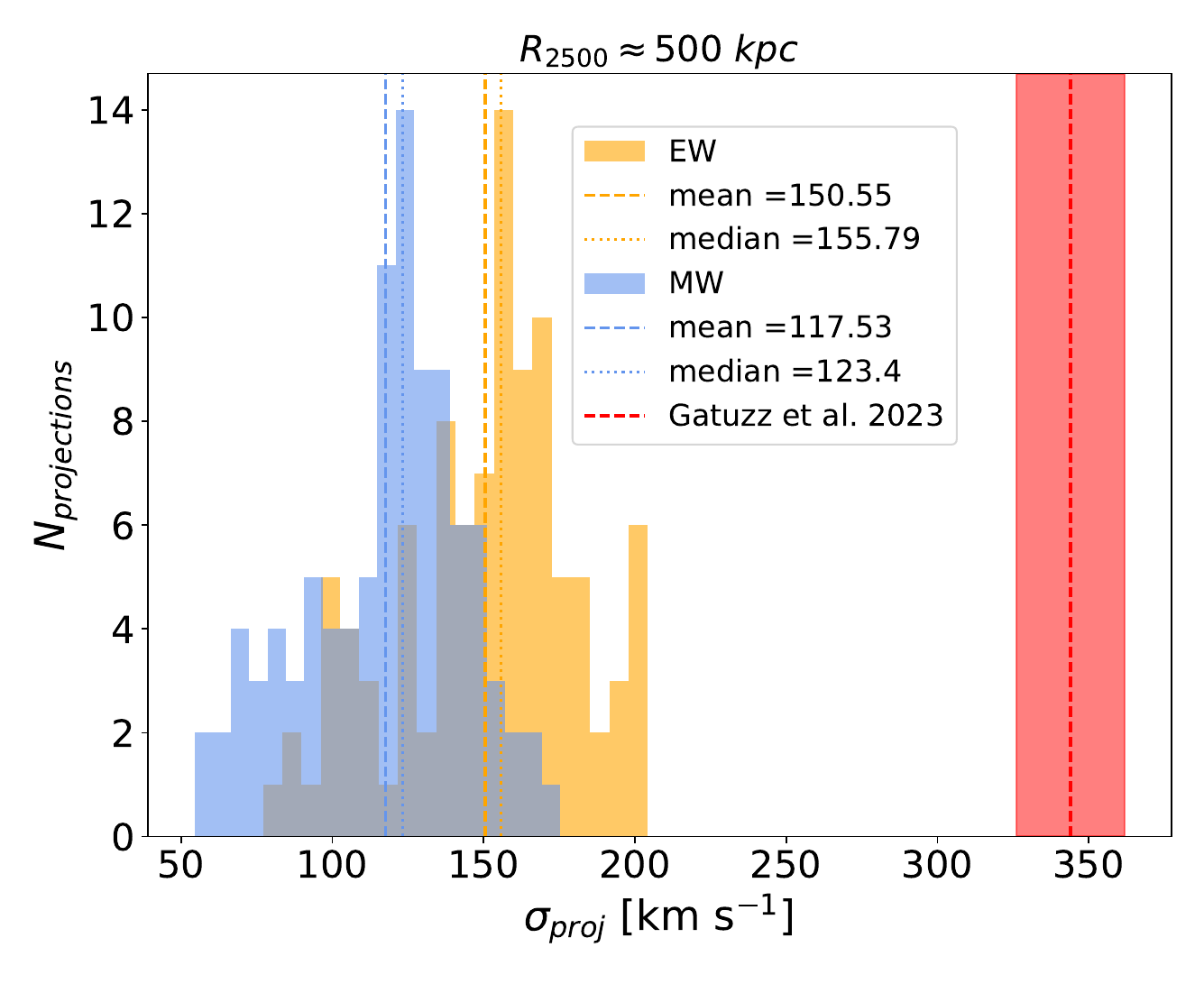}
    \caption{Comparison between $\sigma_{proj}$ of the hundred randomly oriented projections within $R_{\mathrm{2500}}$, as already presented in top left panel of Fig. \ref{sigmas} this time not divided by $\sigma_{sim}$, and the best-fit value of the standard deviation found in \citet{gatuzz2023measuring} displayed as a red dashed vertical line within a shaded area representing the uncertainty.}
    \label{fig:std_comp_gatuzz}
\end{figure}

The VSF is increasingly being used to quantify turbulence in the ICM. On the simulation side, for example \citet{mohapatra2022characterizing} and \citet{2025A&A...698A.121Fournier}, the VSF was computed in the core of clusters, accounting for multiple phases of the gas, magnetic fields, and sub-grid turbulence driving method for the former. They find that the different phases of the gas seem to have uncorrelated VSFs; we did not explore this question in our work since we focused on the hot phase of the gas. In addition, by comparing 3D VSFs to mock observations, \citet{2025A&A...698A.121Fournier} also found that projection effects can reduce the amplitude and thus induce a slightly steeper VSF slope. In addition to effects induced by physics processes in the ICM, projection effects could partially contribute to explaining the steep VSF slope we observed in our work. This could be tested by comparing 3D VSFs to VSFs computed from mock observations, which will be estimated in a future project. Finally, \citet{2025ApJ...984...22Li} compared the VSF of [OII] emission lines emitted by warm gas of central galaxies of four clusters to that of cold molecular gas probed by ALMA CO emission and to Chandra X-ray surface brightness fluctuations. They find a relatively good agreement between the VSF of the three phases, particularly between the cold and warm phases, although they mention that their conclusions have to be taken with caution since systematics and projection effects may bias the results. In this work, we purposely focused on the hot phase of the gas, but the resolution of our simulation, down to 0.3~kpc in the most resolved regions, is sufficient to conduct a study focusing on Virgo's core and explore its different phases to compare to the aforementioned studies, which we might do in the future.\\

Our work is complementary to previous studies of turbulence in Virgo since it focuses on the hot phase of the ICM at larger scales and up to $R_\mathrm{vir}$. For instance, \citet{li2020direct} computed the VSF of $H\alpha$ filaments, i.e. the cold phase of the ICM, using optical MUSE data, and \cite{2018ApJ...865...53Zhuravleva} analysed X-ray surface brightness fluctuations with Chandra. Both these studies focused on the core of Virgo; we thus cannot directly compare their works to ours. \\

Recently, \citet{gatuzz2023measuring} computed the VSF of the ICM of the Virgo cluster from XMM-Newton observations. However, in their study, they computed the VSF at much smaller scales because the observation was centred on the core of Virgo to obtain a sufficiently high signal-to-noise ratio. They computed the VSF in the [3,180]~kpc range, observing a flattening on an $\sim$15~kpc scale, which they consider as a possible driving scale, and a slope of roughly $2/3$ at smaller separations. Given that we computed the VSF on scales larger than 22~kpc and considered the range of validity to be [100,4000]~kpc, we can hardly compare our results to theirs. Still, it might be worth noting that we observe a similar $2/3$ slope in the [100,600]~kpc range and that, given that we observe a potential injection scale at about 600~kpc, we would rather expect a dissipation scale below 20~kpc than another injection scale of the hot phase of the ICM.\\ 

In addition, \citet{gatuzz2023measuring} studied the statistical properties of the velocity field in the core of Virgo. Due to the limited spectral resolution of the EPIC-pn instrument, they were not able to access the sightline velocity dispersion; this will be possible with XRISM and NewAthena. Nonetheless, they estimated the standard deviation of the observed sightline velocities by fitting their distributions to a Gaussian law. For Virgo, they found a standard deviation of 344$\pm$18~$\mathrm{km~s^{-1}}$, which is much higher than any of our results, particularly the standard deviation of the $v_\mathrm{cen}$ projection that is 157.8$\mathrm{km~s^{-1}}$ at maximum. Even the standard deviation of the $x$ component of the 3D velocity field within $R_{vir}$, which is the highest value that we showed to be contaminated by accretion from filaments, has a lower standard deviation of 325.6~$\mathrm{km~s^{-1}}$.  Figure \ref{fig:std_comp_gatuzz} compares the value of their best-fit standard deviation to the distribution of the standard deviation for the 100 randomly oriented projections of sightline velocity within $R_{\mathrm{2500}}\approx500$~kpc. It shows that their value is incompatible with our results, even more so given that their estimation comes from a very central region of the cluster, i.e. less than $(100~\mathrm{kpc})^2$, and that we observe a decreasing standard deviation when decreasing the radius in which it is estimated in our simulation.\\

Before concluding, we summarise the key properties of the Virgo replica simulation impacting the results and the tests conducted on modelling aspects, in particular weighting methods. We also discuss the limitations and possible improvements of this study. First, on the simulation itself, this Virgo replica simulation has the great advantage of being a constrained simulation, thus reproducing the Virgo cluster with great fidelity; this includes its local environment, which we found to play a major role in the gas dynamics and to induce important projection effects. Moreover, this simulation contains a refined AGN feedback model in addition to standard sub-grid models, thus properly accounting for non-gravitational processes, but without implementing a sub-grid model of turbulence \citep[as in e.g.][]{2017MNRAS.469.3641I_Iapichino}. Finally, this Virgo replica's finest AMR cell resolution is 0.35~kpc, thus permitting to develop turbulence down to scales that cannot currently be reached by X-ray observatories. To get closer to scales reachable by current facilities, we produced projections with a resolution of 22.5~kpc (corresponding to a coarser level of refinement of the AMR grid), thus using physical quantities averaged on those scales. This choice indeed reduces the spatial resolution and prevents the study of small-scale features, but, on the other hand, it uses reliable averaged quantities and allows much quicker computation; in particular for the VSF for which we showed (see right panel of Fig. \ref{app: val range fig + vsf comp 222 444}) that a twice as resolved projection yields a very similar result. Then, we discussed the impact of the weighting schemes for the projections, showing that MW projections give results closer to our ground-truth 3D velocity field. We also tested the impact of the weighting scheme of the VSF, using either a constant- or density-weighted scheme (see Fig. \ref{app: VSF weighting compar}), showing similar results except for the Cen projection that is strongly impacted by projection effects. We also found that the 2D VSF does not follow the Kolmogorov prediction, which is certainly due to projection effects and needs to be investigated further. However, two limitations of our study are that we did not estimate the uncertainties on the measured velocity dispersion, as tackled in, for example, \citet{roncarelli2018athena} and in the \citet{2019A&A...629A.143Clerc,2019A&A...629A.144Cucchetti,2024A&A...686A..41Beaumont,2025A&A...702A.215Molin} series of papers, and we did not account for all the instrumental effects. We plan to extend this study in the near future by producing mock observations of XRISM Resolve or New-Athena X-IFU, in which the instrumental effects and uncertainties will be accounted for.

\section{Conclusion}
\label{conclusion}

In this work, we studied the velocity field in the ICM of the Virgo cluster's simulated replica. We first characterised the statistical properties of the velocity field by computing the PDF and the statistical moments of its components from both 3D simulation outputs and projected velocity maps, first in four sightlines and then among 100 random projections. Using the velocity dispersion, we then estimated the non-thermal pressure fraction through an effective turbulent Mach number. Finally, we computed the VSF of the projected velocity maps. We also discussed the results found using either MW or EW projections and the difference in velocity-dispersion estimation between the sightline velocity dispersion and the standard deviation over the sightline velocity in light of other studies.\\

The analysis of the PDFs and the statistical moments shows that the 3D velocity field is not isotropic, particularly in the outskirts, where cosmic gas is accreted at high velocities. It is also not homogeneous as the PDFs tighten up, so the standard deviation reduces when reducing the radius of the sphere from which the gas cells are extracted. Choosing to study the velocity field within $R_\mathrm{500}$ seems to be a good compromise as it is a large enough radius to be statistically representative of the global velocity field, and thus it is neither biased by homogeneities nor too contaminated by the contribution of gas accreted in the outskirts. The projected velocity fields are subject to important projection effects along the $x$ and $cen$ sightlines, given that the latter are quite aligned with filaments connected to Virgo and matter funnelling into it. Overall, the 2D PDFs have smaller dispersions compared to their 3D counterpart, and the MW method shrinks the distributions even more as it smooths the distribution, whereas the EW method highlights the densest regions, leading to more contrasted projections. Generalising over a hundred random projections, we confirmed that the standard deviation of sightline velocity is much lower than its 3D counterpart, but the sightline velocity dispersions, particularly the MW ones, are, on the contrary, in good agreement. \\

We then estimated the non-thermal pressure fraction using the velocity dispersion included in an effective turbulent Mach number. We find a non-thermal pressure fraction of about $6\%$ within $R_\mathrm{500}$ and $9\%$ within $R_\mathrm{vir}$ from direct 3D simulation outputs. The non-thermal pressure fractions estimated using MW sightline velocity dispersion are in good agreement, whereas all the other methods yield a lower value. We also notice the impact of important projection effects for the $x$ component of the 3D velocity field and its associated projection, leading to an overestimation of this fraction by several percent.\\

The first and second-order VSFs of the projected velocity maps have similar slopes below 600~kpc for both EW and MW maps. This range could be considered as the inertial range of the turbulent cascade, so the injection scale, and thus the diameter of the largest eddy, which could be related to the size of the AGN feedback shockfront. Beyond this scale, the VSF shows the contribution of the local large-scale structure, as it increases significantly with separation for the $v_\mathrm{x}$ projected map but more or less flattens for the others. Moreover, we found much steeper slopes than Kolmogorov's predictions, at all scales, indicating that the turbulence in the ICM of this cluster is much more complex than the ideal case. \\

To conclude, the case study of the velocity field in the ICM of the Virgo replica shows how complex its gas dynamics is. It is highly turbulent, and both filaments in the outskirts and the AGN in the core bring energy to the medium and induce a consequential amount of non-thermal pressure. We postpone a systematic study of the impact of this on the reconstruction of the hydrostatic mass to a forthcoming work. Moreover, constraining the properties of the ICM from observations with XRISM and NewAthena will improve our understanding of velocities within the ICM, but there is still a lot to explore on the simulation, theory, and mock observation side to be able to properly analyse the upcoming data. 

\begin{acknowledgements}
The authors acknowledge the Gauss Centre for Supercomputing e.V. (www.gauss-centre.eu) for providing computing time on the GCS Supercomputers SuperMUC at LRZ Munich. This work was supported by the grant agreements ANR-21-CE31-0019 / 490702358 from the French Agence Nationale de la Recherche / DFG for the LOCALIZATION project. This work has been supported as part of France 2030 program ANR-11-IDEX-0003. SE acknowledges the financial contribution from the contracts Prin-MUR 2022 supported by Next Generation EU (n.20227RNLY3 {\it The concordance cosmological model: stress-tests with galaxy clusters}),
ASI-INAF Athena 2019-27-HH.0, ``Attivit\`a di Studio per la comunit\`a scientifica di Astrofisica delle Alte Energie e Fisica Astroparticellare'' (Accordo Attuativo ASI-INAF n. 2017-14-H.0), from the European Union’s Horizon 2020 Programme under the AHEAD2020 project (grant agreement n. 871158), and the support by the Jean D'Alembert fellowship program. The authors thank the IAS Cosmology team members and Nicholas Battaglia for the discussions and comments. The authors thank Florent Renaud for sharing the {\tt rdramses} {\tt RAMSES} data reduction code.
   
\end{acknowledgements}

\bibliographystyle{aa} 
\bibliography{bibliography}

\appendix

\section{Sightline velocity dispersion projections}
\label{app: vel disp los projs}

\begin{figure}[h!]
    \centering
    \includegraphics[trim= 0 0 50 0,clip, width=0.99\linewidth]{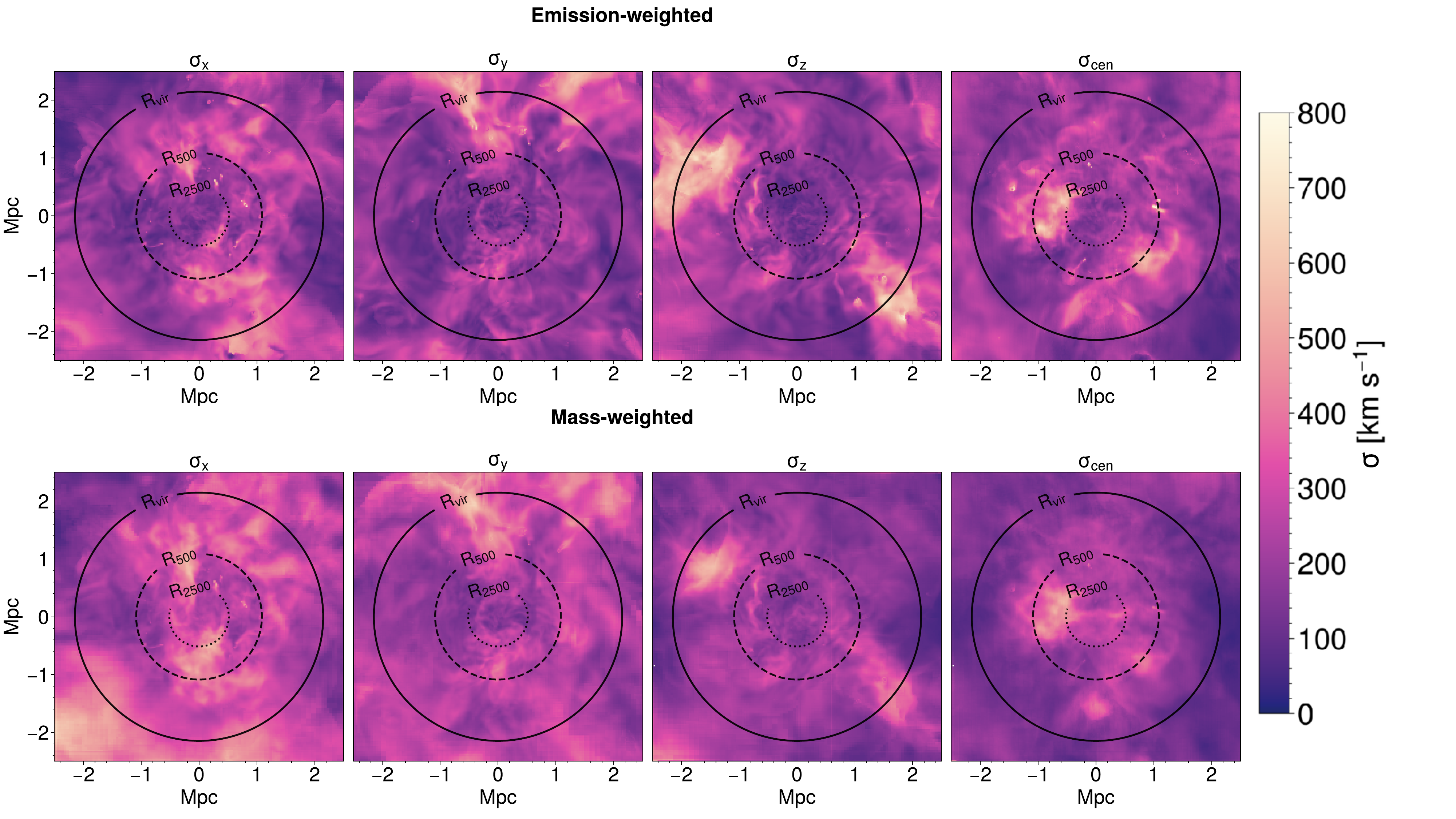}
    \caption{Projected velocity dispersion integrated along the $v_\mathrm{x}$,$v_\mathrm{y}$,$v_\mathrm{z}$ and $v_\mathrm{cen}$ sightlines from left to right. The top (bottom) row presents the EW (MW) projections. The solid, dashed and dotted black circles respectively represent $R_\mathrm{vir}$, $R_\mathrm{500}$ and $R_\mathrm{2500}$.}
    \label{los vel disp maps}
\end{figure}

In Fig. \ref{los vel disp maps}, we show the sightline velocity dispersion alike the sightline velocity presented in Fig. \ref{los vel maps}. The projections seem more contrasted with the emission weighting (top), similarly to the sightline velocity. Moreover, the peaks of velocity dispersion are associated with peaks of sightline velocity, both positive and negative, which shows that although the mean integrated velocity is dominated by the extreme values, there are also smaller velocities contributing along the sightline, which is expected.

\section{Complementary VSF figures}
\label{app:vsf add figs}

\begin{figure}[h!]
    \centering
    \includegraphics[width=0.429\linewidth]{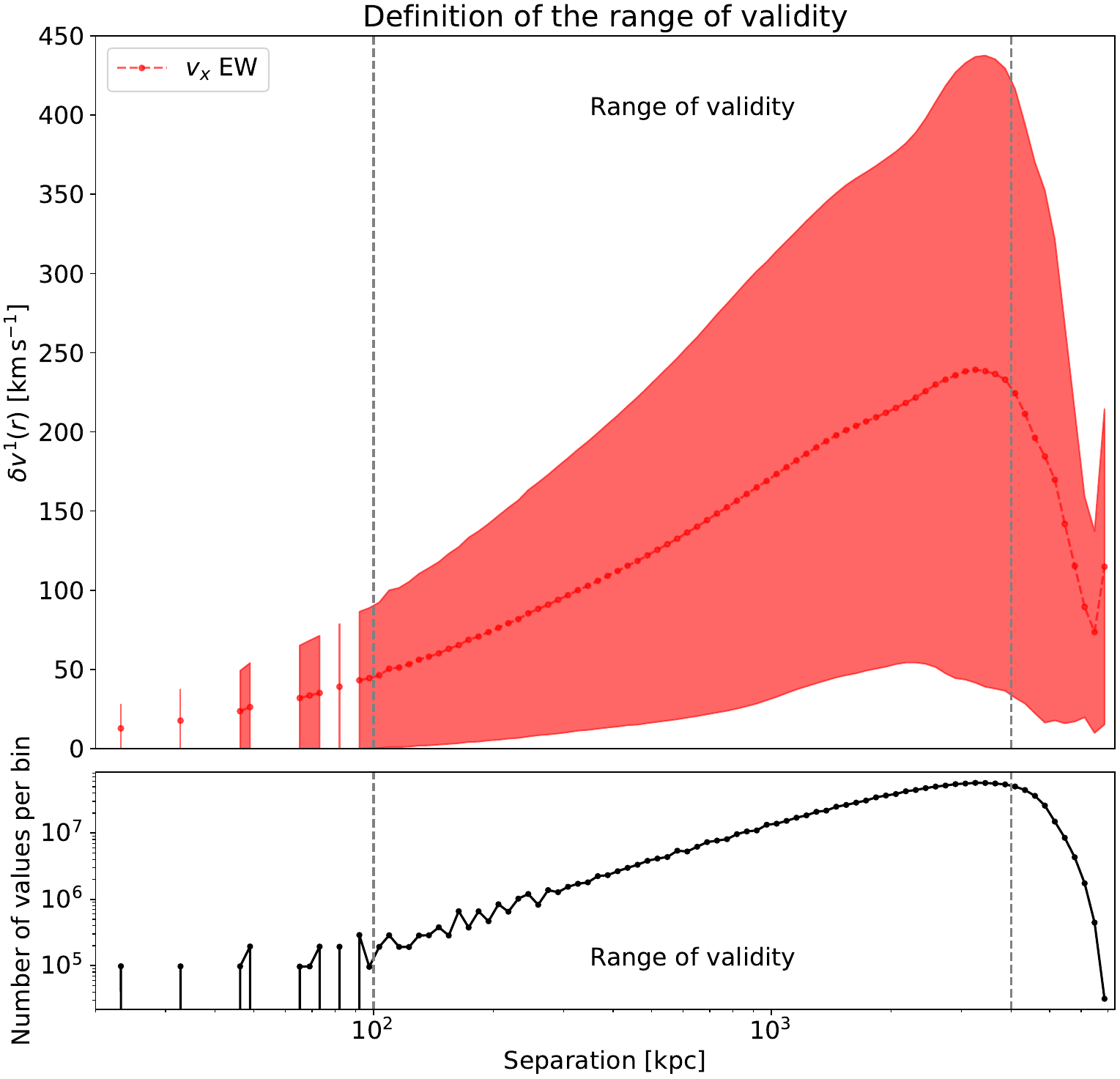}
    \hspace{0.1cm}
    \includegraphics[width=0.502\linewidth]{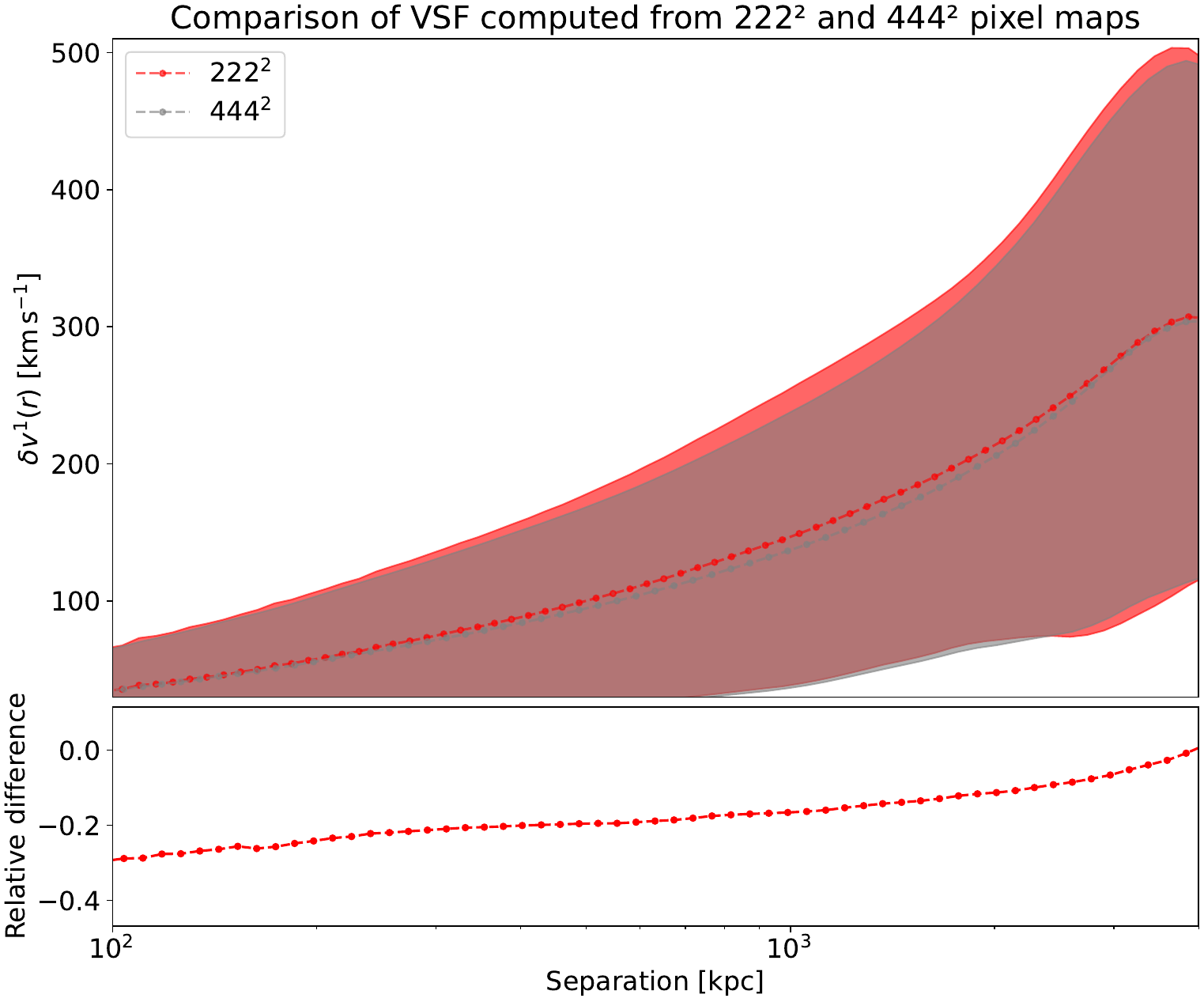}
    \caption{Left: Definition of the range of validity of the VSF, [0.1,4]~Mpc, given the number of values per bin shown on the bottom sub-panel. Right: Comparison of the VSF computed from $222^2$ and $444^2$ pixel maps. The relative difference is shown on the bottom sub-panel.}
    \label{app: val range fig + vsf comp 222 444}
\end{figure}

\begin{figure}
    \centering
    \includegraphics[width=0.99\linewidth]{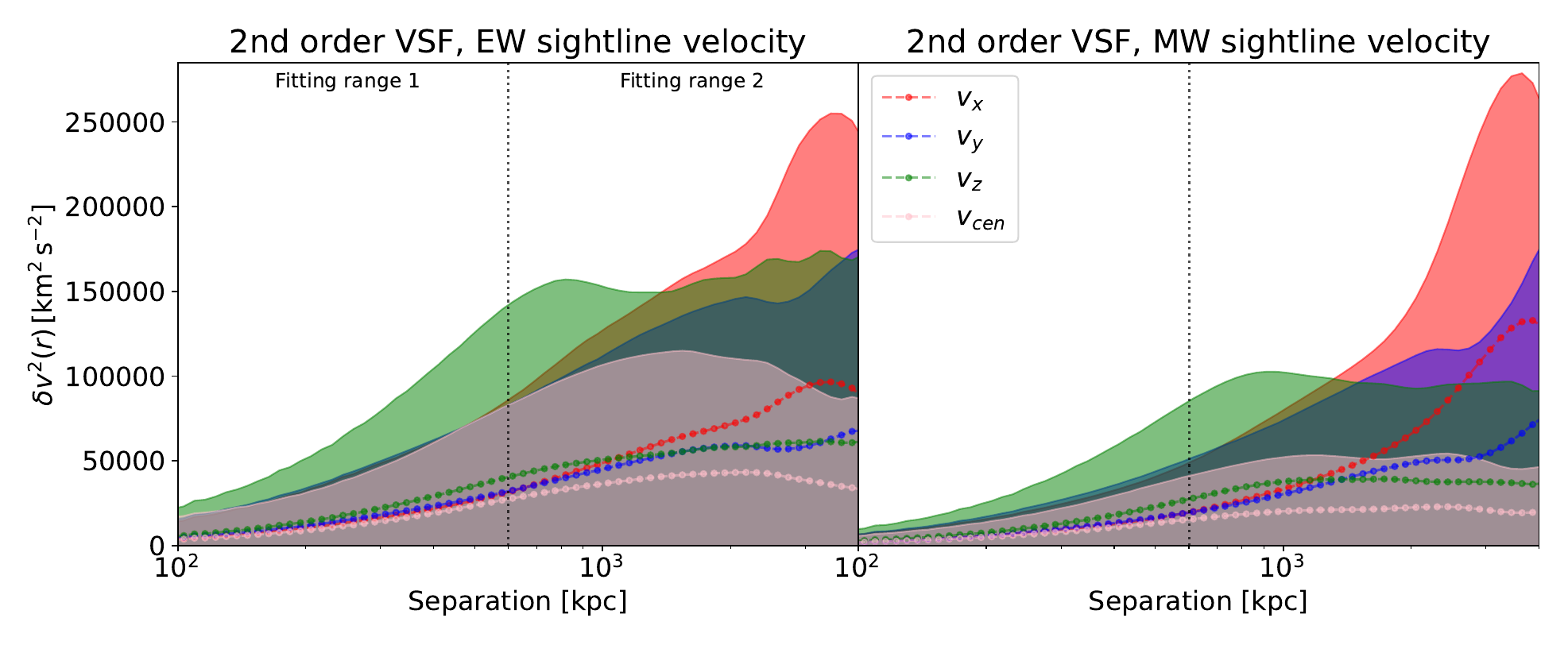}
    \caption{Second order velocity structure function (VSF) computed from EW (left) and MW projected maps (right). They are displayed in red for the $v_\mathrm{x}$ projection, blue for the $v_\mathrm{y}$ projection, green for the $v_\mathrm{z}$ projection and pink for the $v_\mathrm{cen}$ projection. The dispersions are displayed in shaded areas in the same colours. They are presented in the [100,4000]~kpc range, which is the range of validity of the study given the low number of pairs in bins with smaller and larger separations (see left panel of Fig. \ref{app: val range fig + vsf comp 222 444}). The range is separated into two fitting ranges: [100,600]~kpc and [600,1000]~kpc; the delimitation between the two intervals is highlighted by a dashed vertical dotted black line. The best-fit slope values in each interval are reported in Table \ref{tab:fit}.}
    \label{VSF2}
\end{figure}

\begin{figure}
    \centering
    \includegraphics[width=0.8\linewidth]{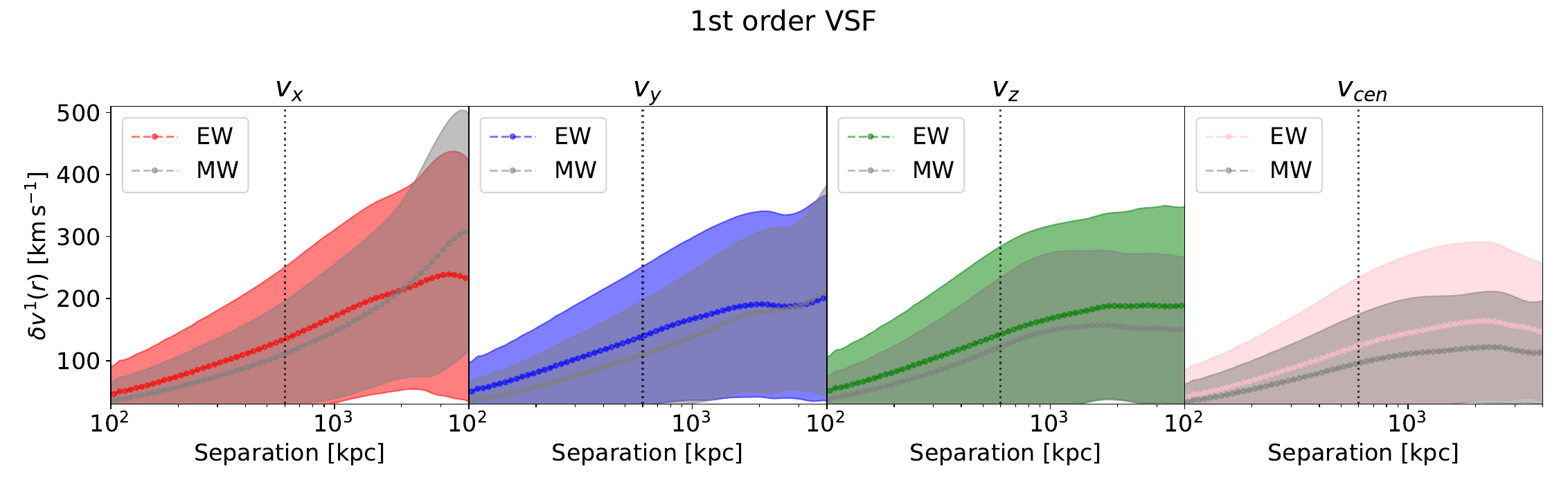}
    \hspace{0.1cm}
    \includegraphics[width=0.8\linewidth]{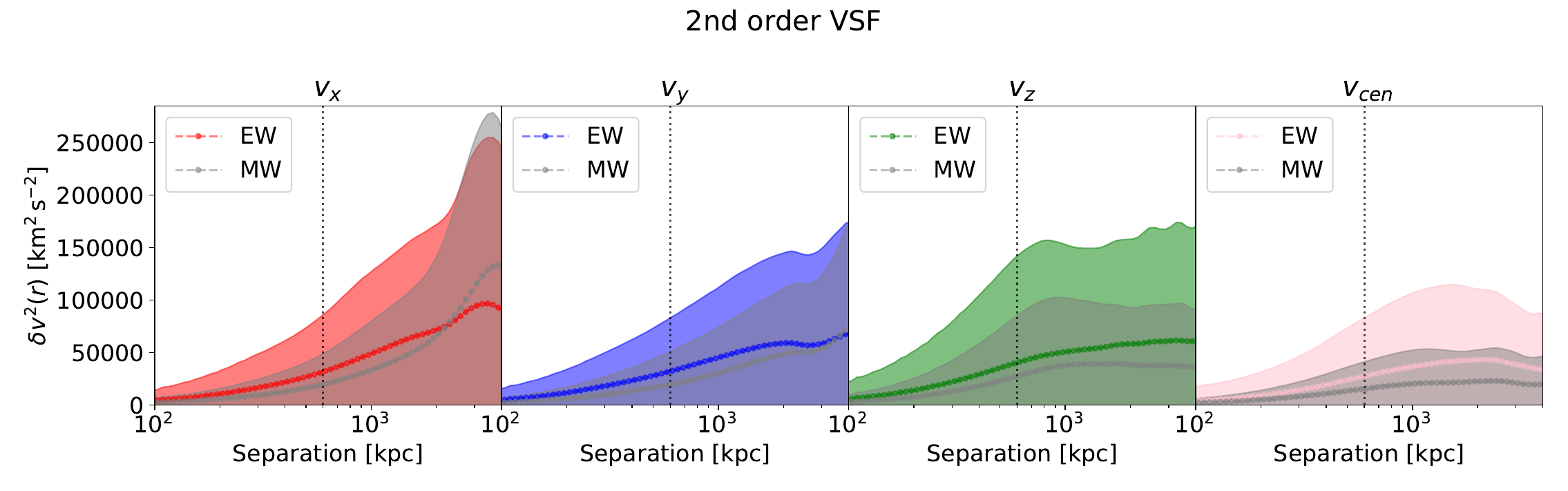}
    \caption{Comparison of the EW and MW (grey) VSFs at the first (top) and second order (bottom).}
    \label{app: VSF EW MW comp}
\end{figure}

\begin{figure}
    \centering
    \includegraphics[width=0.8\linewidth]{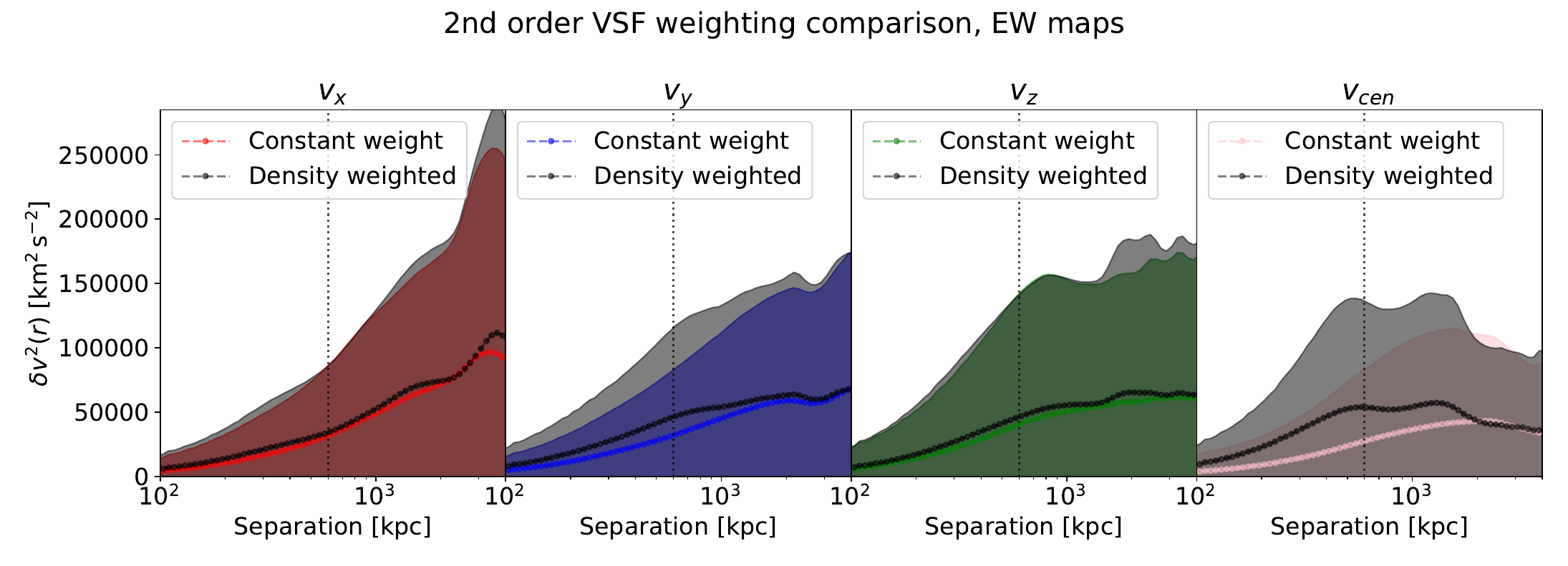}
    \hspace{0.1cm}
    \includegraphics[width=0.8\linewidth]{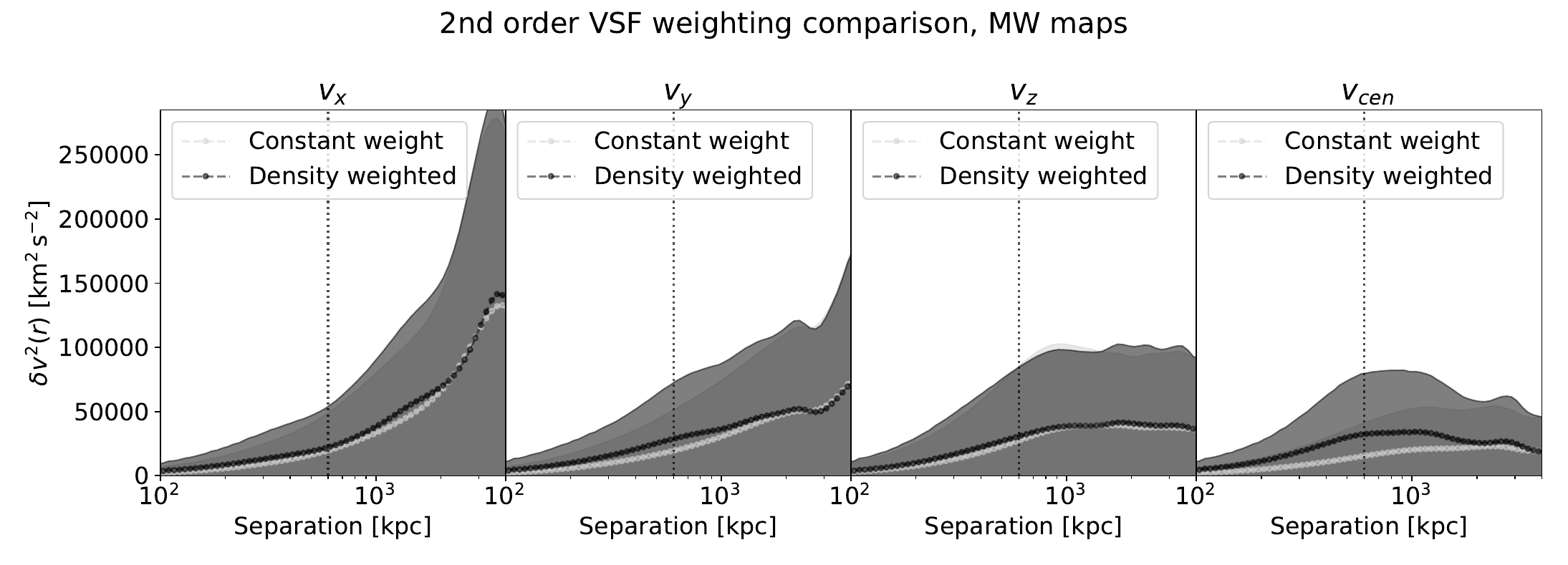}
    \caption{Comparison of the weighting methods used to compute the VSF. The second-order VSF is presented for EW (top) and MW (bottom). We compare the constant-weighted VSF, represented by coloured lines on the top panel and grey lines on the bottom panel, to the density-weighted VSF, displayed in black.}
    \label{app: VSF weighting compar}
\end{figure}

In this section of the Appendix, we show figures completing Sect. \ref{sec:VSF}. First, on Fig. \ref{app: val range fig + vsf comp 222 444}, we show the full range of scales of the VSF on the left panel, and we compare the VSF computed on maps with the resolution used for this study ($222^2$) and twice this resolution ($444^2$) on the right panel, showing an excellent agreement. Second, in Fig. \ref{VSF2}, we show the second-order VSF that is discussed in the core of the article. In Fig. \ref{app: VSF EW MW comp}, we compare the first (top) and second (bottom) VSFs computed on EW (colored) and MW (grey) projections. Finally, in Fig. \ref{app: VSF weighting compar}, we compare two weighting methods to compute the VSF, either constant- or density-weighted \citep[used in e.g.][]{2024A&A...690A..20Ayromlou}. The n-th order density-weighted VSF writes as follows \citep[see e.g.][]{2019A&A...630A..97Chira}:

\begin{equation}
    \delta v^n(\mathbf{r}) = \frac{\sum_\mathrm{i=1}^N \rho(\mathbf{x_i})\rho(\mathbf{x_i}+\mathbf{r})|v(\mathbf{x_i}+\mathbf{r})-v(\mathbf{x_i})|^\mathbf{n}}{\rho(\mathbf{x_i})\rho(\mathbf{x_i}+\mathbf{r})}
\end{equation}

with $\rho$ the gas density. In our case, we used the projected electron column density as already used in \citep[][]{2024A&A...682A.157Lebeau}. We can see that the density-weighted VSF, whose 2nd order is displayed in black in Fig. \ref{app: VSF weighting compar}, shows a very similar amplitude, though slightly higher, and slope compared to the constant-weighted VSF, which is displayed in colours for the EW case (top) and in grey for the MW case (bottom). The only exception is the Cen projection (right panels) for which the amplitude is much higher and the slope steeper for r<600~kpc, though still in the same range of amplitude and slope as the other projections' VSF. This result can be explained by the fact that (i) the sightline velocity is close to zero or even negative in the bottom part of the core (see right panels of Fig. \ref{los vel maps}), the pairs including these pixels have thus an higher absolute velocity difference and (ii) that because the core of the ICM is the densest region, the pairs including pixels in this region core have much more weight than other pairs, explaining why the average velocity difference rises for r<600~kpc. It shows that, in the case of a very strong projection effect due to a filament, its impact can be partly mitigated by computing the density-weighted VSF instead of the constant-weighted since it gives more weight to the velocity in the ICM than in the outskirts. 

\onecolumn

\newpage

\section{Velocity fields statistics}
\label{app:vel stats}

We provide the values of the statistical properties of the PDFs presented in Sect. \ref{sec:3 stat}.

\begin{table}
    \centering
    \scriptsize
    \caption{Statistical properties of the velocity fields.}
    \resizebox{\textwidth}{!}{\begin{tabular}{ c c c c c c c c c }
    \hline \hline \hline
    & \multicolumn{6}{c}{Simulation dataset statistics} & \multicolumn{2}{c}{PDF Gaussian fit best parameters}\\
    & & median & $\mu$ & $\sigma$ & skewness & kurtosis & $\mu$ & $\sigma$ \\
    \hline \hline 
    \multicolumn{9}{c}{3D velocity field properties within spheres}\\
    \hline
    \multirow{3}{*}{$R<R_\mathrm{vir}$} & $v_\mathrm{x}$ &  -37.2 & -9.8 & 325.6 & 0.34 & 2.16 & -44.3 ± 11.1 & 265.0 ± 11.1 \\
    & $v_\mathrm{y}$ & 20.1 & -20.5 & 263.4 & 0.12 & 0.74 & 17.6 ± 5.4 & 243.2 ± 5.4\\
    & $v_\mathrm{z}$ & -30.5 & -30.3 & 260.7 & 0.41 & 0.83 & -38.6 ± 6.9 & 266.2 ± 6.9 \\
    \hdashline
    \multirow{3}{*}{$R<R_{500}$}& $v_\mathrm{x}$ & 1.4 & 2.4 & 249.7 & 0.44 & 2.58 & -0.3 ± 5.1 & 248.2 ± 5.1 \\
    & $v_\mathrm{y}$ & -26.2 & -57.9 & 249.3 & -0.25 & 0.14 & -19.5 ± 5.0 & 253.7 ± 5.0 \\
    & $v_\mathrm{z}$ & -6.4 & -45.7 & 237.7 & 0.20 & 0.15 & -8.9 ± 2.8 & 252.0 ± 2.8 \\
    \hdashline
    \multirow{3}{*}{$R<R_\mathrm{2500}$}& $v_\mathrm{x}$ & -15.4 & -39.0 & 194.3 & 0.35 & 0.50 & -19.5 ± 9.5 & 221.9 ± 9.6 \\
    & $v_\mathrm{y}$ & -67.7 & -95.2 & 228.0 & -0.11 & -0.28 & -70.7 ± 10.6 & 237.6 ± 10.6 \\
    & $v_\mathrm{z}$ &  -90.0 & -118.3 & 203.0 & 0.31 & 0.25 & -98.2 ± 6.8 & 201.0 ± 6.8 \\
    \hline \hline 
    
    \multicolumn{9}{c}{EW projected velocity properties within circles}\\
    \hline 
    \multirow{3}{*}{$R<R_\mathrm{vir}$} & $v_\mathrm{x}$ & -17.1 & -17.8 & 215.6 & -0.25 & 0.37 & -17.9 ± 8.3 & 199.6 ± 8.3 \\
    & $v_\mathrm{y}$ & 34.9 & 13.2 & 168.7 & -0.26 & 0.70 & 36.5 ± 9.1 & 135.9 ± 9.1\\
    & $v_\mathrm{z}$ & -4.6 & 1.4 & 177.5 & 1.03 & 3.15 & -8.7 ± 5.3 & 153.2 ± 5.3 \\
    & $v_\mathrm{cen}$ & 151.9 & 142.0 & 146.9 & 0.33 & 1.41 & 151.8 ± 7.4 & 129.4 ± 7.4 \\
    \hdashline
    \multirow{3}{*}{$R<R_\mathrm{500}$}& $v_\mathrm{x}$ & 5.4 & 10.9 & 137.2 & 0.09 & -0.55 & 8.9 ± 8.1 & 154.2 ± 8.2 \\
    & $v_\mathrm{y}$ & -35.8 & -39.4 & 164.3 & -0.36 & 0.25 & -27.0 ± 6.1 & 146.7 ± 6.1 \\
    & $v_\mathrm{z}$ & 24.8 & -0.9 & 155.8 & -0.58 & -0.16 & 38.6 ± 9.8 & 135.3 ± 10.1\\
    & $v_\mathrm{cen}$ & 167.6 & 160.5 & 153.6 & 0.21 & 0.14 & 160.2 ± 8.5 & 156.4 ± 8.5 \\
    \hdashline
    \multirow{3}{*}{$R<R_\mathrm{2500}$}& $v_\mathrm{x}$ & -5.7 & -8.7 & 124.7 & -0.08 & -0.36 & -3.9 ± 8.9 & 133.1 ± 9.0 \\
    & $v_\mathrm{y}$ & -100.9 & -94.0 & 170.7 & 0.02 & -0.60 & -94.5 ± 8.3 & 191.0 ± 8.8 \\
    & $v_\mathrm{z}$ & -80.8 & -61.7 & 153.1 & 0.05 & -0.79 & -69.4 ± 18.8 & 179.1 ± 20.6 \\
    & $v_\mathrm{cen}$ & 173.1 & 162.6 & 157.8 & -0.10 & -0.74 & 170.9 ± 14.9 & 179.1 ± 15.5 \\

    \hline \hline 
    \multicolumn{9}{c}{MW projected velocity properties within circles}\\
    \hline 
    \multirow{3}{*}{$R<R_\mathrm{vir}$} & $v_\mathrm{x}$ & -37.3 & -50.8 & 219.4 & -0.23 & -0.55 & -28.0 ± 15.7 & 231.1 ± 16.0 \\
    & $v_\mathrm{y}$ & 18.3 & 2.4 & 152.0 & -0.29 & -0.05 & 23.4 ± 7.3 & 139.3 ± 7.3 \\
    & $v_\mathrm{z}$ & -25.2 & -31.2 & 141.9 & 0.38 & 0.74 & -29.9 ± 9.1 & 131.3 ± 9.1\\
    & $v_\mathrm{cen}$ & 98.6 & 97.7 & 104.7 & 0.10 & -0.05 & 97.5 ± 3.5 & 109.6 ± 3.5 \\
    \hdashline
    \multirow{3}{*}{$R<R_\mathrm{500}$}& $v_\mathrm{x}$ & -25.3 & -20.5 & 127.2 & 0.21 & -0.04 & -28.2 ± 3.0 & 127.3 ± 3.0 \\
    & $v_\mathrm{y}$ & -43.1 & -41.8 & 130.5 & -0.24 & 0.45 & -39.9 ± 5.4 & 109.6 ± 5.4 \\
    & $v_\mathrm{z}$ &  11.3 & -12.2 & 129.1 & -0.71 & 0.01 & 26.5 ± 7.0 & 97.1 ± 7.1 \\
    & $v_\mathrm{cen}$ & 90.1 & 91.9 & 129.3 & 0.10 & -0.52 & 87.3 ± 6.7 & 143.7 ± 6.8 \\
    \hdashline
    \multirow{3}{*}{$R<R_\mathrm{2500}$}& $v_\mathrm{x}$ &  -9.8 & -14.9 & 99.7 & -0.01 & -0.16 & -9.6 ± 7.5 & 105.9 ± 7.6 \\
    & $v_\mathrm{y}$ & -97.0 & -80.1 & 135.8 & 0.25 & -0.55 & -96.1 ± 9.6 & 151.4 ± 10.5 \\
    & $v_\mathrm{z}$ & -57.5 & -46.8 & 126.4 & 0.03 & -0.99 & -49.2 ± 17.7 & 164.3 ± 21.2\\
    & $v_\mathrm{cen}$ & 81.0 & 93.8 & 136.0 & 0.47 & -0.42 & 64.6 ± 10.1 & 143.4 ± 11.1 \\
  
    \hline
    \hline
    \hline
    
    \end{tabular}}
    \label{tab:stats PDFs}
    \tablefoot{From left to right: median, first to fourth moments and Gaussian best-fit values of the PDFs. From top to bottom: statistics of the 3D velocity fields, EW and MW projections.}
\end{table}

\newpage

\section{Non-thermal pressure and Mach number values}
\label{app:Pnt}

We provide the values of the quantities calculated in Sect. \ref{sec:4 Mach} within $R_{500}$ in the top sub-panel and within $R_{vir}$ in the bottom sub-panel. 

\begin{table}
    \scriptsize
    \caption{Values of the temperature, sound speed, velocity, velocity dispersion, Mach number and non-thermal pressure fraction.}
    \setlength{\tabcolsep}{2pt}
    \renewcommand{\arraystretch}{1.5}
    \resizebox{\textwidth}{!}{\begin{tabular}{c c c c c c c c c c c c c c}
    \hline \hline \hline
    \multicolumn{13}{c}{Within $R_\mathrm{500}=1087~kpc$} \\
    \hline
     & \multicolumn{4}{c}{3D} & \multicolumn{8}{c}{2D}\\
     T~[K] & \multicolumn{4}{c}{7.077e+07} & & 5.99e+07 & 6.14e+07 & 6.14e+07 & 5.80e+07 & 5.99e+07 & 6.14e+07 & 6.14e+07 & 5.80e+07  \\
     $c_s$~[km s$^{-1}$] & \multicolumn{4}{c}{1.274e+03} & & 1.172e+03 & 1.186e+03 & 1.186e+03 & 1.153e+03 & 1.172e+03 & 1.186e+03 & 1.186e+03 & 1.153e+03 \\
     \hdashline
     & $\overline{v}$ & $\overline{v_\mathrm{x}}$ & $\overline{v_\mathrm{y}}$ & $\overline{v_\mathrm{z}}$ & & $\overline{v_\mathrm{x}}$ & $\overline{v_\mathrm{y}}$ & $\overline{v_\mathrm{z}}$ & $\overline{v_\mathrm{cen}}$ & $\overline{v_\mathrm{x}}$ & $\overline{v_\mathrm{y}}$ & $\overline{v_\mathrm{z}}$ & $\overline{v_\mathrm{cen}}$ \\
     \multirow{2}{*}{$v$~[km s$^{-1}$]} & \multirow{2}{*}{7.39e+01} & \multirow{2}{*}{2.36e+00} & \multirow{2}{*}{-5.79e+01} & \multirow{2}{*}{-4.58e+01} & ew & 1.09e+01 & -3.94e+01 & -8.51e-01 & 1.61e+02 & 1.09e+01 & -3.94e+01 & -8.51e-01 & 1.61e+02 \\
      & & & & & mw & -2.05e+01 & -4.18e+01 & -1.22e+01 & 9.19e+01 & -2.05e+01 & -4.18e+01 & -1.22e+01 & 9.19e+01 \\
      \hdashline
      & $\sqrt{3}\sigma_\mathrm{v}$ & $\sqrt{3}\sigma_\mathrm{v_x}$ & $\sqrt{3}\sigma_\mathrm{v_y}$ & $\sqrt{3}\sigma_\mathrm{v_z}$ & & $\sqrt{3} \sigma_\mathrm{x}$ & $\sqrt{3} \sigma_\mathrm{y}$ & $\sqrt{3} \sigma_\mathrm{z}$ & $\sqrt{3} \sigma_\mathrm{cen}$ & $\sqrt{3} std(v_\mathrm{{}x})$ & $\sqrt{3} std(v_\mathrm{y})$ & $\sqrt{3} std(v_\mathrm{z})$ & $\sqrt{3} std(v_\mathrm{cen})$ \\
      \multirow{2}{*}{$\sigma$ [km s$^{-1}$]} & \multirow{2}{*}{4.25e+02} & \multirow{2}{*}{4.32e+02} & \multirow{2}{*}{4.32e+02} & \multirow{2}{*}{4.12e+02} & ew & 3.78e+02 & 3.15e+02 & 3.09e+02 & 4.54e+02 & 2.38e+02 & 2.85e+02 & 2.70e+02 & 2.66e+02 \\
      & & & & & mw & 5.17e+02 & 3.88e+02 & 3.66e+02 & 4.73e+02 & 2.20e+02 & 2.26e+02 & 2.24e+02 & 2.24e+02 \\
      \hdashline
      \multirow{2}{*}{$M_\mathrm{3D}$} & \multirow{2}{*}{3.39e-01} & \multirow{2}{*}{3.39e-01} & \multirow{2}{*}{3.42e-01} & \multirow{2}{*}{3.25e-01} & ew & 3.22e-01 & 2.68e-01 & 2.61e-01 & 4.17e-01 & 2.03e-01 & 2.42e-01 & 2.28e-01 & 2.69e-01 \\
      & & & & & mw & 4.41e-01 & 3.29e-01 & 3.09e-01 & 4.18e-01 & 1.89e-01 & 1.94e-01 & 1.89e-01 & 2.10e-01 \\
      \hdashline
     \multirow{2}{*}{$\frac{P_\mathrm{nt}}{P_\mathrm{tot}}$} & \multirow{2}{*}{6.00e-02} & \multirow{2}{*}{6.02e-02} & \multirow{2}{*}{6.10e-02} & \multirow{2}{*}{5.55e-02} & ew & 5.46e-02 & 3.83e-02 & 3.63e-02 & 8.82e-02 & 2.24e-02 & 3.16e-02 & 2.80e-02 & 3.87e-02 \\
      & & & & & mw & 9.75e-02 & 5.67e-02 & 5.04e-02 & 8.85e-02 & 1.94e-02 & 2.04e-02 & 1.94e-02 & 2.39e-02 \\      
    \hline
    \hline
    \hline
    \multicolumn{13}{c}{Within $R_\mathrm{vir}=2147~kpc$} \\
    \hline
     & \multicolumn{4}{c}{3D} & \multicolumn{8}{c}{2D} \\
    T~[K] & \multicolumn{4}{c}{5.755e+07} & & 4.04e+07 & 4.16e+07 & 4.15e+07 & 3.98e+07 & 4.04e+07 & 4.16e+07 & 4.15e+07 & 3.98e+07  \\
     $c_s$~[km s$^{-1}$] & \multicolumn{4}{c}{1.149e+03} & & 9.624e+02 & 9.771e+02 & 9.759e+02 & 9.553e+02 & 9.624e+02 & 9.771e+02 & 9.759e+02 & 9.553e+02 \\
     \hdashline
     & $\overline{v}$ & $\overline{v_\mathrm{x}}$ & $\overline{v_\mathrm{y}}$ & $\overline{v_\mathrm{z}}$ & & $\overline{v_\mathrm{x}}$ & $\overline{v_\mathrm{y}}$ & $\overline{v_\mathrm{z}}$ & $\overline{v_\mathrm{cen}}$ & $\overline{v_\mathrm{x}}$ & $\overline{v_\mathrm{y}}$ & $\overline{v_\mathrm{z}}$ & $\overline{v_\mathrm{cen}}$ \\
     \multirow{2}{*}{$v$~[km s$^{-1}$]} & \multirow{2}{*}{3.79e+01} & \multirow{2}{*}{-9.77e+00} & \multirow{2}{*}{-2.05e+01} & \multirow{2}{*}{-3.03e+01} & ew & -1.78e+01 & 1.32e+01 & 1.44e+00 & 1.42e+02 & -1.78e+01 & 1.32e+01 & 1.44e+00 & 1.42e+02 \\
      & & & & & mw & -5.08e+01 & 2.45e+00 & -3.12e+01 & 9.77e+01 & -5.08e+01 & 2.45e+00 & -3.12e+01 & 9.77e+01 \\
      \hdashline
      & $\sqrt{3}\sigma_{v}$ & $\sqrt{3}\sigma_{v_\mathrm{x}}$ & $\sqrt{3}\sigma_{v_\mathrm{y}}$ & $\sqrt{3}\sigma_{v_\mathrm{z}}$ & & $\sqrt{3} \sigma_\mathrm{x}$ & $\sqrt{3} \sigma_\mathrm{y}$ & $\sqrt{3} \sigma_\mathrm{z}$ & $\sqrt{3} \sigma_\mathrm{cen}$ & $\sqrt{3} std(v_\mathrm{x})$ & $\sqrt{3} std(v_\mathrm{y})$ & $\sqrt{3} std(v_\mathrm{z})$ & $\sqrt{3} std(v_\mathrm{cen})$ \\
      \multirow{2}{*}{$\sigma$ [km s$^{-1}$]} & \multirow{2}{*}{4.93e+02} & \multirow{2}{*}{5.64e+02} & \multirow{2}{*}{4.56e+02} & \multirow{2}{*}{4.52e+02} & ew & 3.96e+02 & 3.33e+02 & 3.89e+02 & 4.07e+02 & 3.73e+02 & 2.92e+02 & 3.07e+02 & 2.54e+02 \\
      & & & & & mw & 5.06e+02 & 4.31e+02 & 3.74e+02 & 3.47e+02 & 3.80e+02 & 2.63e+02 & 2.46e+02 & 1.81e+02 \\
      \hdashline
      \multirow{2}{*}{$M_\mathrm{3D}$} & \multirow{2}{*}{4.31e-01} & \multirow{2}{*}{4.91e-01} & \multirow{2}{*}{3.98e-01} & \multirow{2}{*}{3.94e-01} & ew & 4.12e-01 & 3.41e-01 & 3.99e-01 & 4.51e-01 & 3.88e-01 & 2.99e-01 & 3.15e-01 & 3.05e-01 \\
      & & & & & mw & 5.29e-01 & 4.41e-01 & 3.84e-01 & 3.77e-01 & 3.98e-01 & 2.70e-01 & 2.54e-01 & 2.16e-01 \\
      \hdashline
     \multirow{2}{*}{$\frac{P_\mathrm{nt}}{P_\mathrm{tot}}$} & \multirow{2}{*}{9.34e-02} & \multirow{2}{*}{1.18e-01} & \multirow{2}{*}{8.07e-02} & \multirow{2}{*}{7.94e-02} & ew & 8.62e-02 & 6.06e-02 & 8.12e-02 & 1.02e-01 & 7.73e-02 & 4.74e-02 & 5.22e-02 & 4.91e-02 \\
      & & & & & mw & 1.34e-01 & 9.76e-02 & 7.59e-02 & 7.33e-02 & 8.10e-02 & 3.88e-02 & 3.46e-02 & 2.52e-02 \\ 
    \hline
    \hline
    \hline
    
    \end{tabular}}
    \label{app: pnt_mach tab}
    \tablefoot{The top part of the table shows the values calculated within $R_\mathrm{500}$ and the bottom part shows the values calculated within $R_\mathrm{vir}$. To avoid confusion, the standard deviation over the sightline maps is written std.}
\end{table}

\end{document}